\documentclass[a4paper,11pt]{article}
\usepackage{jcappub} 
\usepackage{orcidlink}
\usepackage{multicol}
\usepackage[T1]{fontenc}

\arxivnumber{2502.03573} 
\title{\boldmath Observable Primordial Gravitational Waves from Non-minimally Coupled $R^2$ Palatini Modified Gravity}

\author[a]{Hardik J. Kuralkar\orcidlink{0009-0002-1868-9507},}
\author[a]{Sukanta Panda\orcidlink{0000-0002-1740-5018},}
\author[a]{and Archit Vidyarthi\orcidlink{0000-0003-4269-7128}}

\affiliation[a]{Department of Physics, Indian Institute of Science Education and Research (IISER) Bhopal,\\
Bhopal Bypass Road, Bhauri, Bhopal - 462066, Madhya Pradesh, India}

\emailAdd{kuralkar20@iiserb.ac.in}
\emailAdd{sukanta@iiserb.ac.in}
\emailAdd{archit17@iiserb.ac.in}

\abstract{We probe the spectrum of primordial gravitational waves (GWs) produced during the eras of hyperkination, kination, and reheating in a non-minimally coupled, $\mathcal{L} \propto (1+ \xi \chi /M_{\text{Pl}})^t (R+\alpha R^2)$, modified gravity model using the Palatini formulation under a runaway inflaton potential. The coupling order $t$ is varied to examine a large class of theories up to $\chi^2 R^2$. For models with $t>0$, reheating is not achieved naturally; hence, we supplement such theories with a reheating mechanism based on the interaction of inflaton and radiation produced at the end of inflation due to cosmological expansion which gives successful radiation domination with $T_{\text{reh}} \sim 10^8$ GeV. We demonstrate that the energy density of the GWs is enhanced as a function of the coupling during kination for all considered theories, and a short-lived phase of hyperkination, a result of $R^2$ in Palatini, truncates the boost and avoids the over-production of GWs. Our work provides another strong hint towards inclusion of $R^2$ terms. The spectrum remains flat for the period of hyperkination and reheating. We find that as we decrease the order of the coupling, the spectra shift towards a more observable regime of future GW experiments. We also highlight the observability of GWs from minimal Starobinsky model and availability of wide parameter space. The observation of the plateau during reheating will constrain the $H$ and $\Omega_r^{\text{end}}$ values, while the spectral shape of the boost obtained during kination will confirm the nature of the theory. The bounds from hyperkination lie in the kHz-GHz frequency range. If observed, it will confirm an important prediction of inflation and the existence of a modified theory of gravity.}

\begin{document}
\maketitle
\flushbottom

\section{Introduction}

Cosmic inflation \cite{Albrecht:1982wi,Guth:1980zm,Kazanas:1980tx,Linde:1981mu,Linde:1983gd,Sato:1981qmu,Starobinsky:1980te} stands as the most widely accepted theory providing necessary initial conditions for the Big Bang to occur \cite{Linde_2005,Dodelson_20,Rubakov_Gorbunov_2017,Kolb_2018}. This theory elegantly bypasses the fine-tuning and horizon problem, which has plagued cosmology for a long. It also explained the need for primordial density fluctuations that formed the large-scale structures that we observe today \cite{Starobinsky:1979ty,Mukhanov:1981xt,Hawking:1982cz,Guth:1982ec}. It made predictions about production of the cosmic microwave background (CMB) \cite{Collaboration_Ade_Ahmed_Amiri_Barkats_Thakur_Beck_Bischoff_Bock_Boenish_2021,Collaboration_Akrami_Arroja_Ashdown_Aumont_Baccigalupi_Ballardini_Banday_Barreiro_Bartolo_2019a,Collaboration_Akrami_Arroja_Ashdown_Aumont_Baccigalupi_Ballardini_Banday_Barreiro_Bartolo_2019b}, large-scale structures (LSS) \cite{Linde_2005,Dodelson_20,Mukhanov_2005,Baumann_2022,Gorbunov_Rubakov_2011} and primordial gravitational waves (GWs) from fluctuations \cite{Sahni_1990,Allen_1988,Starobinsky:1979ty}. Observations from WMAP, Planck satellites, and other missions rightfully confirmed the presence of CMB and LSS. Incorporating this phase of early universe accelerated expansion with late time expansion due to dark energy, which the $\Lambda$CDM model explains, gives us the standard model of cosmology.

\begin{figure}
    \centering
    \includegraphics[width=\linewidth]{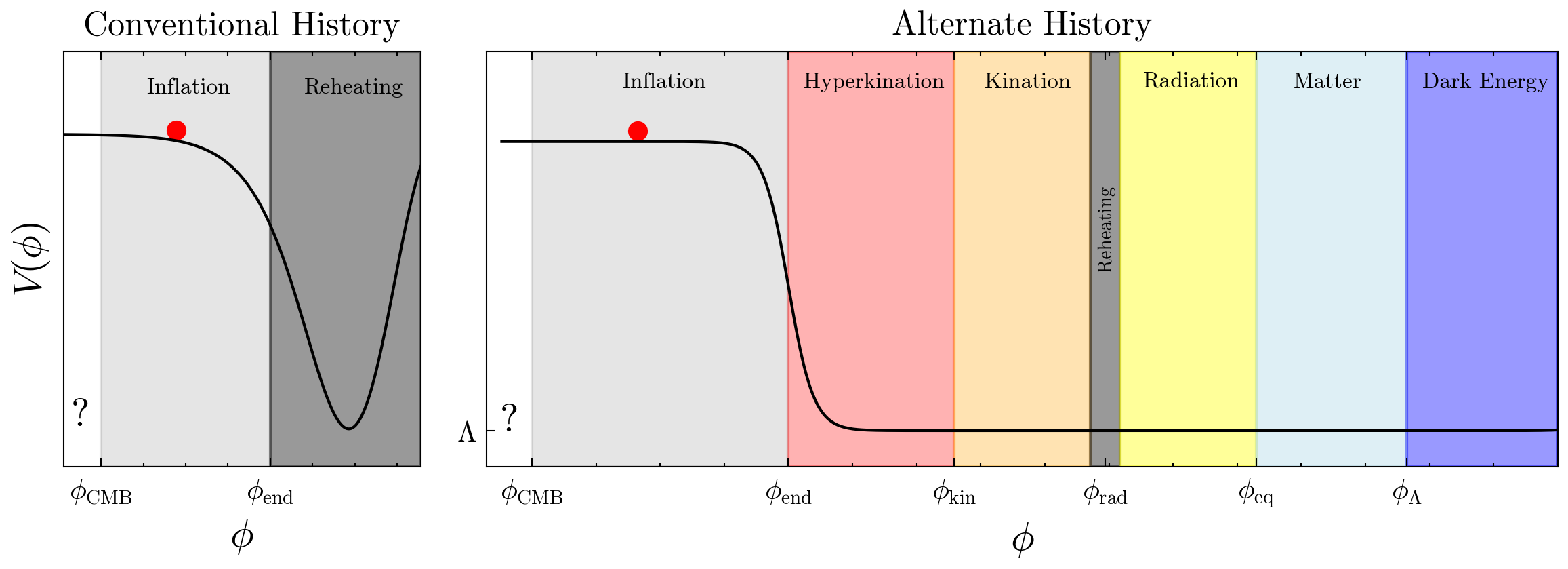}
    \caption{Evolution of the inflaton potential as a function of a field value in conventional (Left) and alternate (Right) history models. $\phi_{\text{CMB}}$ marks the CMB emission, $\phi_{\text{end}}$ the end of inflation and start of hyperkination, $\phi_{\text{kin}}$ the end of hyperkination and start of kination, $\phi_{\text{rad}}$ the end of kination and start of radiation domination, $\phi_{\text{eq}}$ denotes the radiation-matter equality and the onset of matter domination while $\phi_{\Lambda}$ highlights the start of an era dominated by dark energy.}
    \label{fig:cosmic-evol}
\end{figure}

The simplest model of inflation involves a scalar field, also called the \textit{inflaton} field, minimally coupled to gravity, along with a canonical kinetic term and governed by a potential. Traditionally, inflationary theory dictates that the inflaton slowly rolls down a flat and low potential, giving us the required inflation, and then oscillates in a potential minimum and subsequently decays into particles to reheat the universe. However, observation does not support the presence of a potential minimum, unlike the plateau that gives inflation.  As an alternative, models with an inflaton potential that do not have a minimum (or a region where the scalar field cannot oscillate \cite{Felder_Kofman_Linde_1999}) after the inflationary plateau and where the scalar field freely rolls down under its kinetic energy were considered (see Figure \ref{fig:cosmic-evol}). As the scalar field comes down from the potential dominated inflation, it freely rolls, gaining kinetic energy. This period of kinetic domination is called \textit{kination} \cite{Gomez_Lola_Pallis_Rodriguez-Quintero_2009,Joyce_Prokopec_1997,Pallis_2005,Pallis_2006}. Such models, called quintessential inflation models \cite{Peebles_Vilenkin_1998,Peloso_Rosati_1999,Huey_Lidsey_2001,Dimopoulos_2000}, try to explain the late-time expansion of the universe due to dark energy and early universe inflation in a unified theory. Supplemented with observational constraints from \cite{Collaboration_Ade_Aghanim_Arnaud_Ashdown_Aumont_Baccigalupi_Banday_Barreiro_Bartlett_2016,Collaboration_Aghanim_Akrami_Ashdown_Aumont_Ballardini_Banday_Barreiro_Bartolo_Basak_2017,Array_Collaborations_Ade_Ahmed_Aikin_Alexander_Barkats_Benton_Bischoff_Bock_2016} require that the inflaton potential to be very flat at large field values, which can be achieved by a non-minimal coupling between inflaton field and gravity or introducing quadratic or higher order curvature terms in the action \cite{Collaboration_Ade_Ahmed_Amiri_Barkats_Thakur_Beck_Bischoff_Bock_Boenish_2021}. However, this was not the only motivation that demanded the presence of these additional terms.

Almost half a decade after Einstein's seminal work on the general theory of relativity (GR), people found that it is not renormalizable. It was shown that a renormalizable Einstein-Hilbert action demands higher-order curvature terms \citep{Utiyama_1962}, and these higher-order actions are indeed renormalizable \citep{Stelle:1976gc}. These claims have also been supported by string-theoretic calculations \citep{Vilkovisky:1992pb,Buchbinder_1989,Birrell:1982ix}. This prompted the scientific community to explore higher-order gravitational theories, namely \textit{modifications} of the Einstein–Hilbert action to incorporate higher-order curvature invariants related to the Ricci scalar. 

Another motivation comes from astrophysical and cosmological observations. No known theoretical models have been able to explain the observations. The simplest model that currently fits the data is the concordance model, $\Lambda$CDM. To name a few, issues like the presence of dark energy, cosmic acceleration, the existence of dark matter, and the large-scale structure of the universe came up from observations that were not explained by GR. The search for theories that explain these observations and also meet theoretical expectations led to the rise of the domain of \textit{modified gravity}. 
One such class of theory that strikes out is the $f(R)$ theories of gravity \citep{De_Felice_2010, Sotiriou_2010,Nojiri_2011,Nojiri_2017}. It simply generalizes the Einstein-Hilbert action as a function of the Ricci scalar,
\begin{equation}
    \textbf{S} = \frac{1}{2\kappa}\int \mathrm{d}^4 x \sqrt{-g} ~f(R) 
\end{equation}
where $f(R)$ can take the form of any polynomial function. Apart from their motivation from high-energy physics, this class of theories is simple enough to handle and work with. These theories also avoid the fatal Ostrogradski instability \citep{Woodard_2007} that makes it special among other higher-order theories.

The choice of variation principle becomes essential when working with such a class of theories. The \textit{metric variation} is the standard one where we vary the action with respect to the metric. The second, popularly known as \textit{Palatini variation} \cite{Palatini:1919ffw}, is where the metric and the connection are considered independent of each other and the action is varied with respect to both. However, the Palatini variation assumes that the action for matter does not depend on the connection. Both formalisms result in the same set of equations for the Einstein-Hilbert action and actions, which are linear in $R$. However, results differ when higher-order terms are included or when considering a more general action (see \cite{Gialamas_2023} and references therein). As fewer assumptions about the relationship between gravitational degrees of freedom are considered in the Palatini, it is more natural to work in it. In addition, the Palatini formulation of Higgs inflation preserves tree-level unitarity better than its metric counterpart, making the model better behaved from a quantum field-theoretic point of view \cite{Shaposhnikov_Shkerin_Zell_2021,Panda_2023,Mikura_Tada_2022,McDonald_2021,Ito_Khater_Rasanen_2022,Bauer_Demir_2011,Antoniadis_Guillen_Tamvakis_2021}.

One of the key signatures from inflation and the period after it are primordial GWs, \textit{primordial} because they are \textit{old}. These are a direct result of the accelerated expansion of the universe and are produced as a consequence of tensor perturbations of the spacetime metric and constitute a stochastic gravitational-wave background (SGWB). Primordial GWs were considered widely in modified gravity theories \cite{Odintsov_2021,Odintsov_2022,Capozziello_2017}. An indirect artifact of these GWs is the imprint on CMB in the form of B-mode polarizations, which can be observed \cite{Turner_1997}. While these GWs can also be observed directly through interferometers, like the LIGO (Laser Interferometer Gravitational-Wave Observatory) \cite{aLIGO_2010,aLIGO_2015}, Virgo \cite{aVIRGO_2015}, and KAGRA \cite{KAGRA_2021}, and experiments like the NANOGrav (North American Nanohertz Observatory for Gravitational Waves) \cite{NANOGrav_2013,NANOGrav_2019}, PTAs (Pulsar Timing Arrays) \cite{PTA_1990} like the European PTA (EPTA) \cite{EPTA,EPTA_2008}, the International PTA (IPTA) \cite{IPTA_2016,IPTA_2019}, and the Parkes PTA (PPTA) \cite{PPTA_2013}, they are too faint to detect with our current detection capabilities. However, with future experiments like LISA (Laser Interferometer Space Antenna) \cite{LISA_2016,LISA_2017,LISA_2019,LISA_2023}, ET (Einstein Telescope) \cite{ET_2010}, CE (Cosmic Explorer) \cite{CE_2019}, BBO (Big Bang Observatory) \cite{BBO_2004,BBO_2005,BBO_2006}, SKA (Square Kilometer Array) \cite{SKA_2009}, DECIGO (DECi-hertz Interferometer Gravitational-wave Observatory) \cite{DECIGO_2006,DECIGO_2011,DECIGO_2021}, ASTROD-GW (Astrodynamical Space Test of Relativity using Optical Devices optimized for Gravitational Wave) \cite{ASTRODGW_2013}, $\mu$Ares \cite{muAres}, AION-Km (Atom Interferometer Observatory and Network) \cite{AION} and AEDGE (Atomic Experiment for Dark Matter and Gravity Exploration) \cite{AEDGE} such observations can be positively anticipated. Detection of ultra-high frequency GWs through resonant cavities also shows promise \cite{Aggarwal_2021}. If observed, this will allow us to probe the early universe at very high energies and beyond the last scattering surface. 

In this paper, we try to probe the spectral energy density today of the primordial gravitational waves generated during the epochs of hyperkination, kination, and reheating. Working in the Palatini formalism, we examine a modified gravity of the form $\mathcal{L} \propto (1+ \xi \chi /M_{\text{Pl}})^t (R + \alpha R^2)$, where $\chi$ is our inflaton field, $R$ is the Ricci scalar, and $t$ is the variable order of our coupling. $R^2$ models in Palatini formalism have been widely studied in the context of inflation and reheating \cite{Gialamas:2019nly,Gialamas:2020snr,Gialamas:2021enw,Gialamas:2022xtt,Gialamas:2024jeb}. The order parameter $t$ let us probe lower-order theories ($t<0$) \cite{Kaneda_2016}, the Starobinsky model ($t=0$) \cite{Starobinsky:1980te} up to $\chi^2 R^2$ ($t=2$). We work with a non-oscillatory, runaway inflaton potential  (see Figure \ref{fig:cosmic-evol}) that allows the field to roll down freely after inflation. Starting with the modified gravity action in the Jordan frame, we move to the Einstein frame through a Weyl transformation to obtain non-canonical quartic kinetic terms. The quartic kinetic terms are a unique signature of $R^2$ and are responsible for the period of \textit{hyperkination} \cite{Sanchez_Lopez_2023}, which can occur just after inflation and before kination. We then solve for the dynamics of the field during the epochs of kination and hyperkination. To counter the issue of reheating in this model with non-oscillatory potential, we supplement a reheating mechanism. The radiation produced at the end of inflation modifies the Lagrangian by coupling to the inflaton field, causing decay of scalar field energy density and bringing out radiation domination and, thus, a successful reheating. We calculate the tensor perturbations and their spectrum analytically during all phases of hyperkination, kination, reheating, and beyond. We examine the available parameter space for which the theories remain valid and place bounds on the Hubble parameter ($H$) and radiation energy density ($\Omega_r^{\text{end}}$) at the end of inflation. We explore the observability of these spectra against the sensitivity of the current and future GW detection experiments. A successful detection, if realized in the future, will strongly hint towards the presence of higher-order curvature terms and a modified theory of gravity.

The paper is organized as follows: In Section \ref{sec:model}, we discuss the mathematical formulation of the model with an overview of the action, equation of motion, and the periods of inflation, kination, and hyperkination. In Section \ref{sec:reheating}, we propose a reheating mechanism that supplements our model. We provide an insight into background solutions and gravitational wave mode functions and compute their spectrum in Section \ref{sec:GW}, while Section \ref{sec:GWObs} contains our results on the observability of primordial gravitational waves by current and future experiments. We conclude the paper with a brief summary in Section \ref{sec:conclusion}. We provide details of our calculations and derivations in the appendices.

For this paper, we work in natural units $c = \hbar = 1$ and $8\pi G = M_{\text{Pl}}^{-2}$ where $M_{\text{Pl}} = 2.43 \times 10^{18}$ GeV is the reduced Planck mass. The signature of our metric will be mostly positive.

\section{The Model} \label{sec:model}

\subsection{The Action and the Potential}

We start with a Jordan frame action in the Palatini formalism with a non-minimal and very general coupling of the scalar field with gravity and a higher-order curvature term,
\begin{align} \label{eq:action_01}
    \textbf{S} 
    &= \int \mathrm{d}^4 x \sqrt{-g} \left[ \frac{1}{2 \kappa} h (\varphi) \left( R + \alpha R^2 \right)   -\frac{1}{2} g^{\mu \nu} \partial_\mu \varphi \partial_\nu \varphi - V(\varphi) \right] + \textbf{S}_m [g_{\mu \nu}, \psi]
\end{align}
where, $\varphi$ is the inflaton field, $h(\varphi)$ is its non-minimal coupling function. $g = \text{det}(g_{\mu \nu})$, and $\kappa = 8\pi G = M_{\text{Pl}}^{-2}$ is called the Einstein gravitational constant, with $G$ being the Newton's gravitational constant. We can write, $f(R,\varphi) \equiv h (\varphi) \left( R + \alpha R^2 \right)$ where $R$ is the Ricci scalar defined as $R = g^{\alpha \beta} R^\gamma_{\alpha \gamma \beta} (\Gamma, \partial \Gamma)$ and we assume $\alpha$ to be a positive constant. We leave the potential $V(\varphi)$ undefined for now. The second part of the action $\textbf{S}_m [g_{\mu \nu}, \psi]$ houses other matter components of the universe. 

We introduce a new auxiliary field $\phi$ to rewrite the gravitational part in terms of an action linear in $R$ plus a new scalar field. The action thus becomes,
\begin{align} \label{eq:action_02}
    \textbf{S} 
    &= \int \mathrm{d}^4 x \sqrt{-g} \left[ \frac{1}{2 \kappa} f(\phi,\varphi) + \frac{1}{2 \kappa} \tilde{f}(\phi,\varphi) (R-\phi^2) - \frac{1}{2} g^{\mu \nu} \partial_\mu \varphi \partial_\nu \varphi - V(\varphi) \right] + \textbf{S}_m [g_{\mu \nu}, \psi]
\end{align}
where $\tilde{f}$ represents derivative with respect to the auxiliary field $\phi$ and $\tilde{\tilde{f}}$ represents double derivative with respect to $\phi$. Note that the $\phi^2$ field does not have any kinetic term present, so varying the action with $\phi^2$ will not result in any equation of motion but rather a constraint equation.
\begin{equation}
    \delta \textbf{S} = 0
    \implies \phi^2 = R \hspace{0.3cm} \text{for} \hspace{0.3cm} \tilde{\tilde{f}} \neq 0
\end{equation}
Using the above constraint in \eqref{eq:action_02}, we retrieve \eqref{eq:action_01}. A little rearrangement in \eqref{eq:action_02} gives,
\begin{equation} \label{eq:action_05}
    \textbf{S} = \int \mathrm{d}^4 x \sqrt{-g} \left[ \frac{1}{2 \kappa} \tilde{f}(\phi,\varphi)R - \frac{1}{2 \kappa} W(\phi, \varphi) - \frac{1}{2} g^{\mu \nu} \partial_\mu \varphi \partial_\nu \varphi - V(\varphi) \right]
\end{equation}
where, $W(\phi, \varphi) \equiv \phi^2 \tilde{f}(\phi, \varphi) - f(\phi, \varphi)$. We focus on the epoch when the other matter components $\psi$ is a perfect fluid of radiation, which makes the coupling between the inflaton field and matter weak \cite{Dimopoulos_2022}. Thus, it is safe to neglect the last term in action \eqref{eq:action_02}. Here, the action \eqref{eq:action_05} describes a single scalar field $\varphi$, which is non-minimally coupled to gravity. To obtain a minimally coupled scalar field in the Einstein frame, we perform the Weyl transformation. The transformation depends on both $\phi$ and $\varphi$ and can be written as,
\begin{align}
    g_{\mu \nu} & \rightarrow \Omega^2 \Bar{g}_{\mu \nu} = \tilde{f}(\phi, \varphi) \Bar{g}_{\mu \nu}
\end{align}
This gives, 
\begin{equation}
    \sqrt{-g} \rightarrow \tilde{f}(\phi, \varphi)^2 \sqrt{-\Bar{g}} \hspace{0.3cm} \text{and} \hspace{0.3cm} g^{\mu \nu} \rightarrow \frac{1}{\tilde{f}(\phi, \varphi)} \Bar{g}^{\mu \nu}
\end{equation}
We write the transformed action in the Einstein frame as\footnote{We drop the overhead bars to avoid clutter},
\begin{equation} \label{eq:action_03}
    \textbf{S} = \int \mathrm{d}^4 x \sqrt{-g} \left[ \frac{R}{2 \kappa} - \frac{1}{2 \tilde{f}(\phi,\varphi)} g^{\mu \nu} \partial_\mu \varphi \partial_\nu \varphi - \Hat{V}(\phi, \varphi) \right]
\end{equation}
where the conformally transformed potential is defined as,
\begin{equation}
    \Hat{V}(\phi, \varphi) \equiv \frac{1}{\tilde{f}(\phi,\varphi)^2} \left[ \frac{W(\phi, \varphi)}{2 \kappa} + V(\varphi) \right] = \frac{1}{\tilde{f}(\phi,\varphi)^2} \left[ \frac{\alpha }{2 \kappa}  h(\varphi) \phi^2 + V(\varphi) \right]
\end{equation}
Varying the action with respect to $\phi^2$ and equating it to zero gives us a constraint equation that for $\tilde{\tilde{f}} / \tilde{f}^2 \neq 0$  provides us with,
\begin{align}
    \Hat{V} (\phi, \varphi)  & = \frac{1}{h(\varphi) + 8 \alpha \kappa  V(\varphi)} \left[ \frac{\alpha \kappa }{2} (g^{\mu \nu} \partial_\mu \varphi \partial_\nu \varphi)^2 +  \frac{V(\varphi)}{h(\varphi)} \right]
\end{align}
Note that the above potential is devoid of $\phi$. The action in \eqref{eq:action_03} thus becomes,
\begin{align}
    \textbf{S} 
    &= \int \mathrm{d}^4 x \sqrt{-g} \left[ \frac{R}{2 \kappa} - \frac{1}{2} \frac{g^{\mu \nu} \partial_\mu \varphi \partial_\nu \varphi}{h(\varphi) + 8 \alpha \kappa  V(\varphi)} + \frac{\alpha \kappa} {2} \frac{(g^{\mu \nu} \partial_\mu \varphi \partial_\nu \varphi)^2}{h(\varphi) + 8 \alpha \kappa V(\varphi)}   -  \frac{1}{h(\varphi) + 8 \alpha \kappa V(\varphi)}\frac{V(\varphi)}{h(\varphi)}  \right]
\end{align}
The above action is set up in the Einstein frame, and thus, the potential  and the field redefinition in the same to obtain a canonical kinetic term for a new field $\chi$ can be defined as,
\begin{equation} \label{eq:field_redef}
    U(\varphi) \equiv \frac{V(\varphi)}{h(\varphi)(h(\varphi) + 8 \alpha \kappa V(\varphi))} \quad \text{and} \quad  \frac{\mathrm{d} \varphi }{\mathrm{d} \chi} \equiv \pm \sqrt{h(\varphi) + 8 \alpha \kappa V(\varphi)}
\end{equation}
We obtain the final action as,
\begin{equation} \label{eq:action_04}
    \textbf{S}
    = \int \mathrm{d}^4 x \sqrt{-g} \left[ \frac{R}{2 \kappa} - \frac{1}{2} g^{\mu \nu} \partial_\mu \chi \partial_\nu \chi + \frac{\alpha \kappa}{2} (h(\chi) + 8 \alpha \kappa V(\chi)) (g^{\mu \nu} \partial_\mu \chi \partial_\nu \chi)^2   -  U(\chi)  \right] 
\end{equation}
Comparing with \eqref{eq:action_01}, we can see that the $R^2$ term has translated into the quartic kinetic term, and we have a new potential for the scalar field, which is an important feature of our model.

\subsection{Equation of Motion}

Now, we want to find the energy-momentum tensor. We can write \eqref{eq:action_04} as,
\begin{equation}
    \textbf{S} = \int \mathrm{d}^4 x \sqrt{-g} \left[ \frac{R}{2 \kappa} + \mathcal{L}_{\chi}  \right] 
\end{equation}
where,
\begin{equation}
    \mathcal{L}_{\chi} \equiv - \frac{1}{2}g^{\mu \nu} \partial_\mu \chi \partial_\nu \chi + \frac{\alpha \kappa }{2} (h(\chi) + 8 \alpha \kappa V(\chi)) (g^{\mu \nu} \partial_\mu \chi \partial_\nu \chi)^2   -  U(\chi)
\end{equation}
The least-action principle demands that the variation of action with respect to the metric field is zero, $\delta \textbf{S}=0$ and since the variation holds for any $\delta g^{\mu \nu}$, it implies,
\begin{equation} \label{eq:EOM_metric}
    - \frac{1}{4 \kappa} g_{\mu \nu} R +  \frac{1}{2 \kappa} \frac{\delta R}{\delta g^{\mu \nu}} 
    = \frac{1}{2}  g_{\mu \nu} \mathcal{L}_{\chi} -  \frac{\delta \mathcal{L}_{\chi}}{\delta g^{\mu \nu}}
\end{equation}
This is the equation of motion for the metric field. By definition, the RHS of \eqref{eq:EOM_metric} is the stress-energy (energy-momentum) tensor $T_{\mu \nu}$. We use the metric,
\begin{equation}
    g_{\mu \nu} = \text{diag} (-1, a^2,a^2,a^2)
\end{equation}
where, $a(\tau)$ is the scale factor. From the non-zero components of the energy-momentum tensor and working in the unitary gauge, i.e., $\chi = \chi (\tau)$, we get,
\begin{align}
    \begin{split} \label{eq:rho-chi2}
        T_{00}
    &=  \rho_{\chi} =  \frac{1}{2} \left[ 1 + 3 \alpha \kappa (h + 8 \alpha \kappa V) \Dot{\chi}^2 \right]\Dot{\chi}^2 +  U 
    \end{split} \\
    \frac{1}{3 a^2} \sum_i T_{ii}
    &= p_{\chi} = \frac{1}{2} \left[ 1 + \alpha \kappa (h + 8 \alpha \kappa V) \Dot{\chi}^2 \right] \Dot{\chi}^2 -  U 
\end{align}
where the overhead dot represents the derivative with respect to time $\tau$.

In the limit $\alpha \rightarrow 0$, the above two equations reduce to the minimal case. We now move towards getting the dynamical equation for the scalar field $\chi$. Solving the Euler-Lagrange equation for  $\mathcal{L}_\chi' = \sqrt{-g} ~\mathcal{L}_\chi$ we get, 
\begin{align}
    2 \left[1 + 3 \alpha \kappa(h +8 \alpha \kappa V)  \Dot{\chi}^2 \right] \Ddot{\chi} + 3 \left[ 1 + \alpha \kappa (h+8 \alpha \kappa V) \Dot{\chi}^2 \right] 2H \Dot{\chi} + \frac{3 \alpha \kappa}{2} (h' +8 \alpha \kappa V') \Dot{\chi}^4 + U' &=0
\end{align}
where prime denotes derivative with respect to the field $\chi$.

\subsection{Inflation}

For slow-roll inflation, the potential dominates, and the quartic kinetic terms in \eqref{eq:action_04} can be safely neglected. Slow-roll inflation is defined by the slow-roll parameters as,
\begin{equation}
    \epsilon_\text{v} = \frac{1}{2} \left( \frac{U'}{U} \right)^2 = \frac{1}{2} \left( \frac{V' h^2 - 2h' V (h+4 \alpha \kappa V)}{hV(h + 8 \alpha \kappa V)} \right)^2
\end{equation}
\begin{align}
    \begin{split}
        |\eta_\text{v}| &= \frac{|U''|}{U} = \frac{1}{h^2 V (h + 8 \alpha \kappa V)^2} \left[ 128 \alpha^2 \kappa^2 h'^2 V^3 - 2h^3 \left[ 2V' (h'+4 \alpha \kappa V') +V h'' \right] \right.\\
        & \left. + 6h^2 V (h'^2 - 4\alpha \kappa h'' V) + 16 \alpha \kappa h V^2 (3 h'^2 - 4 \alpha \kappa V h'') + h^3 V'' (h + 8 \alpha \kappa V) \right]
    \end{split}
\end{align}
The scalar spectral index and tensor-to-scalar ratio can then be calculated
from the following equations as
\begin{equation}
    r = 16 \epsilon_\text{v} \quad \text{and} \quad n_s = 1 + 2 \eta_\text{v} - 6 \epsilon_\text{v}
\end{equation}
For successful slow-roll inflation, $r \to 0$ and $n_s \to 1$. The viable models that give theoretically accepted values were given by \cite{Das_2021}
\begin{multicols}{2}

    \begin{enumerate}
    \item $V = \beta \chi^4$ and $h = \gamma \chi^2$
    \item $V = \beta \chi^2$ and $h = \frac{1}{2} + \gamma \chi^2$
    \item $V = \beta \chi^4$ and $h = \frac{1}{2} + \gamma \chi^2$
    \end{enumerate}

\end{multicols}
However, we find a model, $V = \beta \chi^2$ and $h = \gamma \chi^2$ (more in Appendix \ref{app:inflation}), as valid and claim that successful inflation is observed in $\chi^2 R^2$ theory with a quadratic potential. In addition to being consistent with Planck observations \cite{Panda_2023b}, the model can also be consistent with sub-Planckian inflation and with potential corrections from quantum gravity whilst conserving unitarity during inflation and reheating to a sufficient temperature for a successful post-inflation cosmology \cite{Lloyd-Stubbs:2020pvx,Panda_2023}. We can assume that inflation is successfully achieved for all orders of couplings with runaway potential as the behavior of the potential is the same for all couplings before the end of inflation.

\subsection{Hyperkination and Kination}

After inflation has ended, the kinetic terms assert their dominance and bring the quartic and quadratic terms into the scenario. In this period of kinetic domination, the scalar field $\chi$ takes over the potential, which becomes negligible and rolls freely. In this limit ($V \rightarrow 0$) we get
\begin{align}
    \begin{split} \label{eq:rho-chi3}
        \rho_{\chi} &=  \frac{1}{2} \left[ 1 + 3 \alpha \kappa h \Dot{\chi}^2 \right]\Dot{\chi}^2
    \end{split} \\
    p_{\chi} &= \frac{1}{2} \left[ 1 + \alpha \kappa h \Dot{\chi}^2 \right] \Dot{\chi}^2 
\end{align}
\begin{equation}
    2 \left[ 1 + 3 \alpha \kappa h  \Dot{\chi}^2 \right] \Ddot{\chi} + 3 \left[ 1 + \alpha \kappa h \Dot{\chi}^2 \right] 2H \Dot{\chi} + \frac{3 \alpha}{2} h' \Dot{\chi}^4  =0
\end{equation}
\begin{equation} \label{eq:ff_kd}
    3H^2 = \kappa \rho_{\chi}
\end{equation}
where $H$ is the Hubble parameter. Following the methodology given by \cite{Sanchez_Lopez_2023}, in order to eliminate $H$ from the equation of motion, we change the time variable to the number of $e$-folds, $N = \ln{a}$, where $\mathrm{d} N = H \mathrm{d} \tau$. We normalize the scale factor such that, for $a = e^N = 1$ for $N=0$, $N$ denotes the number of $e$-folds after the end of inflation. Let the overhead bar denote the derivative with respect to $N$ (the prime on $h$ indicates derivative with respect to $\chi$) and let us assume $\Dot{\chi}>0$. From \eqref{eq:ff_kd},
\begin{equation}
    3 H^2 = \frac{\kappa}{2} \left[ 1 + 3 \alpha \kappa h \Dot{\chi}^2 \right]\Dot{\chi}^2 = \frac{\kappa}{2} \left[ 1 + 3 \alpha \kappa h \Dot{\chi}^2 \right] \left( \frac{H \mathrm{d} \chi}{\mathrm{d} N}\right)^2 
\end{equation}
Solving for $\Dot{\chi}$ gives \footnote{We replace $\kappa$ by $M_{\text{Pl}}^{-2}$ at some places for simplicity.},
\begin{equation} \label{eq:chi_relation}
    \Dot{\chi} = \sqrt{\frac{6 M_{\text{Pl}}^2 - \Bar{\chi}^2}{3 \alpha \kappa h \Bar{\chi}^2}}
\end{equation}
Substituting \eqref{eq:chi_relation}, we obtain a new set of equations,
\begin{align}
    \begin{split} \label{eq:rho-chi}
        \rho_{\chi} &=  \frac{6 M_{\text{Pl}}^2 - \Bar{\chi}^2}{\alpha \kappa h \Bar{\chi}^4} M_{\text{Pl}}^2 
    \end{split}\\
    p_{\chi} &= \frac{(\Bar{\chi}^2 + 3 M_{\text{Pl}}^2) (6 M_{\text{Pl}}^2 - \Bar{\chi}^2)}{9 \alpha \kappa h \Bar{\chi}^4}\\
    w_{\chi} &= \frac{1}{9 M_{\text{Pl}}^2} \left( \Bar{\chi}^2 + 3 M_{\text{Pl}}^2 \right)\\
    \begin{split} \label{eq:chi''}
         \Bar{\Bar{\chi}} & = \frac{\Bar{\chi}}{24 M_{\text{Pl}}^4} (6 M_{\text{Pl}}^2 - \Bar{\chi}^2) \left[ -\frac{h'}{h} \left( 1 + \frac{\Bar{\chi}^2}{6 M_{\text{Pl}}^2} \right) \Bar{\chi} M_{\text{Pl}}^2 + \frac{4}{3} \left( \Bar{\chi}^2 + 3 M_{\text{Pl}}^2 \right) \right]
    \end{split}
\end{align}
where the barotropic parameter of the field, $w_{\chi}$ is defined as,
\begin{equation}
    w_{\chi} \equiv \frac{p_\chi}{\rho_\chi} 
\end{equation}
The derivation for \eqref{eq:chi''} is listed in detail in Appendix \ref{app:chi}. Consider a limit when $\Dot{\chi}$ is very small, i.e. $\Dot{\chi} \rightarrow 0$, then $\Bar{\chi} \rightarrow \sqrt{6} M_{\text{Pl}}$. The above equations give,
\begin{align}
    \Bar{\Bar{\chi}} &\approx \sqrt{6} (\Bar{\chi} - \sqrt{6} M_{\text{Pl}} ) \left[ \frac{h'}{h} M_{\text{Pl}} - \sqrt{6} \right] \implies \Bar{\chi} \approx \sqrt{6} M_{\text{Pl}} (1 \pm c e^{\sqrt{6} dN})\\
    \rho_{\chi} &\approx \frac{\sqrt{6}(\sqrt{6} M_{\text{Pl}} - \Bar{\chi})}{18 \alpha \kappa h M_{\text{Pl}}} \approx \mp \frac{c}{3 \alpha \kappa h} e^{\sqrt{6} dN} \propto a^{ \sqrt{6} d}\\
    w_\chi &\approx 1
\end{align}
where, $d = \left[ \frac{h'}{h} M_{\text{Pl}} - \sqrt{6} \right]$ and $c$ is the constant of integration. 

For $\Bar{\chi} = \sqrt{6} M_{\text{Pl}}$ to be a kinetic attractor in the large $N$ limit, we demand $d<0$. Consider any coupling of the form $h(\chi) = (1+ \xi \chi /M_{\text{Pl}})^t$, where $t \in \mathbb{R}$ and $\xi$ is a dimensionless coupling parameter. This results to $d = t \xi (1+\xi \chi /M_{\text{Pl}})^{-1} - \sqrt{6}$. For $d<0$,
\begin{equation}
    \frac{t \xi }{1+\xi \chi /M_{\text{Pl}}} < \sqrt{6}
\end{equation}
For couplings with the form $t\leq 2$, any value of the field, small or large, will satisfy this bound. This era described by the above limit is called \textit{kination} or \textit{regular/standard kination}. We work in the small field regime, which gives us,
\begin{equation}
    \rho_{\chi}  \propto a^{t\sqrt{6}-6}
\end{equation}

In the limit, when $\Dot{\chi}$ is very large, i.e. $\Dot{\chi} \rightarrow \infty$, then $\Bar{\chi} \rightarrow 0$. This gives,
\begin{align}
    \Bar{\Bar{\chi}} &\approx \Bar{\chi} \implies \Bar{\chi} \approx ce^{N} \propto a\\
    \rho_{\chi} &\approx \frac{6 M_{\text{Pl}}^4}{\alpha \kappa h} \frac{1}{\Bar{\chi}^4} \approx \frac{6c}{\alpha \kappa h} e^{-4N} \propto a^{-4}\\
    w_\chi &\approx \frac{1}{3}
\end{align}
where $c$ is the constant of integration. This limit describes the era of \textit{hyperkination}.

We demand that the transition from hyperkination to kination be continuous,
\begin{equation} \label{eq:transition}
    \Bar{\chi} =\begin{cases}
			\Bar{\chi}_0 e^{N}, & N <  \ln{(\sqrt{6} M_{\text{Pl}} /\Bar{\chi}_0)} \ldots \text{Hyperkination}\\
            \sqrt{6} M_{\text{Pl}}, & N >  \ln{(\sqrt{6} M_{\text{Pl}} /\Bar{\chi}_0)}  \ldots \text{Kination}
		 \end{cases}
\end{equation}
where $\Bar{\chi}_0$ is the initial value of $\Bar{\chi}$ at $N=0$. $N=0$ corresponds to the end of inflation. We can define the length of the hyperkination phase from the above equation as,
\begin{equation} \label{eq:deltaN}
    \Delta N_{\text{hyp}} \equiv \ln{\left( \frac{\sqrt{6} M_{\text{Pl}}}{\Bar{\chi}_0} \right)} \equiv N_{\text{hyp}}^{\text{end}}
\end{equation}

When the potential $V(\phi)$ grows to a very large value, the modified potential in the Einstein frame goes to the asymptotic value of $U \to \frac{1}{8 \alpha \kappa h}$, see \eqref{eq:field_redef}. This bounds the inflationary and post-inflationary energy density with an upper bound.
\begin{equation}
    \rho_\chi < \frac{1}{8 \alpha \kappa h}
\end{equation}
From \eqref{eq:rho-chi},
\begin{equation}
    \frac{6 M_{\text{Pl}}^2 - \Bar{\chi}^2}{\alpha \kappa h \Bar{\chi}^4} < \frac{1}{8 \alpha \kappa h} \implies \Bar{\chi}_0 > 2 M_{\text{Pl}}
\end{equation}
Since the potential decays from a large value, it is sensible that the above restriction gives an estimate of the initial value of $\Bar{\chi}$. From \eqref{eq:deltaN}, $\Delta N_{\text{hyp}} \leq 0.20$.

Consider \eqref{eq:rho-chi2},
\begin{equation}
    \rho_{\chi} =  \frac{1}{2} \Dot{\chi}^2   + \frac{3 \alpha \kappa}{2} (h + 8 \alpha \kappa V) \Dot{\chi}^4 +  U
\end{equation}
During different phases of evolution, different terms in the energy density equation dominate. Inflationary epoch demands the domination of the potential, the third term $U$. The inflationary epoch ends when the potential $U$ equals the sum of the first two terms. After this, the potential $V$ becomes negligible, and kinetic terms take over. $R^2$ models have two kinds of kinetic terms governing the evolution. First, the quartic terms dominate, giving the period of hyperkination, followed by the kination, which happens due to the canonical quadratic kinetic terms. Thus, the beginning of standard kination, $N_{\text{kin}}$, is defined as the moment when both the addends in the parenthesis in \eqref{eq:rho-chi3} become equal. 
\begin{equation}
    1 = 3 \alpha \kappa h \Dot{\chi}^2 \implies 1 = \frac{6 M_{\text{Pl}}^2 - \Bar{\chi}^2}{\Bar{\chi}^2} \implies \Bar{\chi} = \sqrt{3} M_{\text{Pl}}
\end{equation}
From \eqref{eq:transition}, $N_{\text{kin}} = \ln (\sqrt{3} M_{\text{Pl}} /\Bar{\chi}_0)$. Comparing to \eqref{eq:deltaN},
\begin{equation} \label{eq:transition_length}
    N_{\text{hyp}}^{\text{end}} - N_{\text{kin}}  = \ln \sqrt{2} \approx 0.35
\end{equation}
Note that this value does not depend on the initial condition or the form of the coupling. It suggests that kination starts before hyperkination. It can be interpreted as follows: As the potential domination starts to decline, the kinetic energy of the field overtakes, resulting in the start of kinetic domination. As soon as the potential is negligible, the quartic kinetic terms bring their effect, giving hyperkination. This is then followed by kination, where the quadratic terms govern the dynamics.

\begin{figure}
    \centering
    \includegraphics[width=\textwidth]{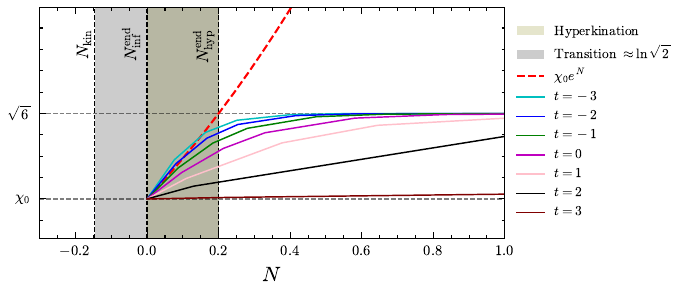}
    \caption{Comparison of the convergence of the derivative of the field with respect to the number of $e$-folds $N$ to the kinetic attractor value (dashed horizontal line) for \eqref{eq:transition} and for different exponents of coupling $t$ from \eqref{eq:chi''}. The three dashed vertical lines denote the start of kination, $N_{\text{kin}}$, the end of inflation, and the start of hyperkination, $N_{\text{inf}}^{{\text{end}}} = 0,$ and the limiting case for the length of hyperkination, $N_{\text{hyp}}^{{\text{end}}}=0.2$. The transition period is highlighted in grey, while hyperkination is in olive green. The epoch of hyperkination is fastest for the initial approximation \eqref{eq:transition}, and the length increases as we increase the coupling exponent in \eqref{eq:chi''}. The field derivative for $t=3$ never approaches the attractor value.}
    \label{fig:phi-evol}
\end{figure}

We compare the convergence of the derivative of the field with respect to the number of $e$-folds $N$ to the kinetic attractor value for \eqref{eq:transition} and for different exponents of coupling $t$ from \eqref{eq:chi''}, considering all the constraints on initial conditions and length of hyperkination and transition in Figure \ref{fig:phi-evol}. The three black dashed vertical lines denote the start of kination, $N_{\text{kin}}$, the end of inflation, the start of hyperkination, $N_{\text{inf}}^{{\text{end}}} = 0,$ and the limiting case for length of hyperkination, $N_{\text{hyp}}^{{\text{end}}}=0.2$. The constraint on transition length between hyperkination and kination from \eqref{eq:transition_length} suggests that the kinetic terms start to dominate slightly before the inflation completely ends and hyperkination lasts for a small duration. The end of hyperkination is followed by the phase of standard kination, where the field rolls forward freely. The epoch of hyperkination is fastest for the initial approximation in \eqref{eq:transition}. The length increases as we increase the coupling exponent in \eqref{eq:chi''}, and convergence is achieved at values mentioned in Table \ref{tab:no_of_efolds}. The field derivative for $t=3$ never approaches the attractor value.

Note that the bound on the $N_{\text{hyp}}=0.2$ comes from the bound on energy density, which cannot be ignored. These lengths in Table \ref{tab:no_of_efolds} are small compared to the required number of $e$-folds for inflation, $\Delta N_{\text{inf}} \sim 50-60$. As we will see in the following sections, the presence of hyperkination is very important, even if it is for a small duration. We consider that kination starts after hyperkination ends for simpler computations. However, we do not ignore the fact that there is a transition period between those phases.

\begin{table}[h]
    \centering
    \begin{tabular}{rccccccc}
    \hline
    \hline
    Coupling & -3 & -2 & -1 & 0 & 1 & 2 & 3\\ \hline
    $N_{\text{hyp}}^{\text{end}}$ & 0.76 & 0.81 & 1.39 & 1.64 & 2.24 & 6.18 & -\\ \hline
    \end{tabular}
    \caption{The values at which the $N$-derivative of the field, solutions to \eqref{eq:chi''}, converges to the attractor value for different couplings $t$. The length of hyperkination decreases as we move to lower-order couplings. For every case, the inflation ends at $N=0$. No convergence is observed for $t=3$.}
    \label{tab:no_of_efolds}
\end{table}

\section{Reheating Mechanism} \label{sec:reheating}

The standard expansion history starts with a scalar field rolling down a potential to a minimum where it oscillates, converting all its potential energy to kinetic energy, thus (re)heating the universe and decaying into particles, providing seeds to the formation of large-scale structures in the universe. However, we considered a runaway potential, where instead of the scalar field getting trapped in a minimum, it rolls down a flat and low potential under kinetic domination. This period of kinetic domination can be branched into those phases, one where the quartic kinetic terms dominate - hyperkination, which later die down and are taken over by the quadratic kinetic terms giving the period of kination.

A small amount of radiation is required after inflation to reheat the universe. During this time, the energy density of the universe is dominated by the scalar field. From the last section, we saw that during hyperkination, the energy density of the scalar field and radiation goes as $\rho_{\chi, \text{r}} \propto a^{-4}$. During the kination era, the field decays as $\rho_\chi \propto a^{t\sqrt{6}-6}$, which if $t > 0$ decays slower than radiation. This does not naturally give us radiation domination from the decay of energy density of the scalar to values lower than that of radiation. To achieve reheating, we provide a mechanism motivated by \cite{Kofman_1994}.

After inflation, the sub-dominant vacuum fluctuations grow large, leading to the creation of particles \cite{Ford_1987}. However, this cosmological gravitational particle production is not sufficient to bring radiation domination and is plagued with issues \cite{Yahiro_Mathews_Ichiki_Kajino_Orito_2002,Riazuelo_Uzan_2000,Giovannini_1999b,Giovannini_1999a, Chun_Scopel_Zaballa_2009,Boyle_Buonanno_2007}. The radiation generated at the end of inflation modifies the Lagrangian by introducing interaction terms between the scalar field and the particles. These interaction terms are further responsible for the decay of the scalar field to those particles. This decay diminishes the energy density of the field, which brings out the required radiation domination and a successful reheating. 

The modified Lagrangian, at the end of inflation, is given as,
\begin{equation}
    \mathcal{L} = \mathcal{L}_\chi + \mathcal{L}_\zeta + \mathcal{L}_\psi + \mathcal{L}_I
\end{equation}
where the first three terms govern the dynamics of the inflaton field $\chi$, bosons $\zeta$, and fermions $\psi$. The final term governs the interaction between those fields and is the one which is responsible for bringing out the decay,
\begin{equation}
    \mathcal{L}_I = -\frac{1}{2}b^2 \chi^2 \zeta^2 - n \Bar{\psi} \psi \chi
\end{equation}
where $b$ and $n$ are small and positive coupling constants. For simplicity, we assume that the bare masses of the fields $\zeta$ and $\psi$ are very small, such that $m_\zeta (\chi) = b \chi$ and $m_\psi (\chi) = |n \chi|$. We can consider the effective potential of the scalar field after kination as,
\begin{equation}
    V(\chi) = \pm \frac{1}{2} m_{\chi}^2 \chi^2 + \frac{\lambda}{4} \chi^4
\end{equation}
where $\lambda$ is a small coupling constant. A classical scalar field $\sigma = m_\chi / \sqrt{\lambda}$ is produced when the scalar field undergoes spontaneous symmetry breaking (SSB) $\chi \to \chi + \sigma$ because of the minus sign. $-\frac{1}{2}b^2 \chi \sigma \zeta^2$ originates from  $-\frac{1}{2}b^2 \zeta^2 \chi^2$ due to SSB. The scalar field can only decay fully if single scalar $\chi$-particle can decay into other particles as a result of the processes $\chi \to \zeta \zeta$ and $\chi \to \Bar{\psi} \psi$. Such processes cannot occur in our model if there is no SSB and no interactions with fermions.

Originally, the decay rate was matched with the decreasing energy of oscillations of the inflaton field with an oscillatory potential. However, as opposed to the elementary theory of reheating \cite{Linde_2005,Abbott_1982,Dolgov_1982}, the nature of the potential is runaway-like in our model. Therefore, we attribute the decay to the expansion of the universe caused by inflation. The rates of the decay processes $\chi \to \zeta \zeta$ and $\chi \to \Bar{\psi} \psi$ with $m_\chi \gg 2 m_\zeta, 2 m_\psi$ are given as,
\begin{equation}
    \Gamma (\chi \to \zeta \zeta) = \frac{b^4 \sigma^2}{8 \pi m_\chi} = \frac{b^4 m_\chi}{8 \pi \lambda}, \quad \Gamma (\chi \to \Bar{\psi} \psi) = \frac{n^2 m_\chi}{8 \pi}
\end{equation}

The total decay rate is given as,
\begin{equation}
    \Gamma = \Gamma (\chi \to \zeta \zeta) + \Gamma (\chi \to \Bar{\psi} \psi) = \left( \frac{b^4}{\lambda} + n^2 \right) \frac{m_\chi}{8 \pi}
\end{equation}

Typically, coupling constants of the interaction of the inflaton field with matter are extremely small \cite{He_2019}, we consider $b \sim \mathcal{O}(0.1)$ and $n \sim \mathcal{O}(0.01)$. $\lambda$ is the coupling constant for the Higgs-like potential. Considering it is stable, we choose $\lambda \sim \mathcal{O}(0.01)$. The only free parameter now is the mass of the scalar field. 

\begin{figure}
    \centering
    \includegraphics[width=\textwidth]{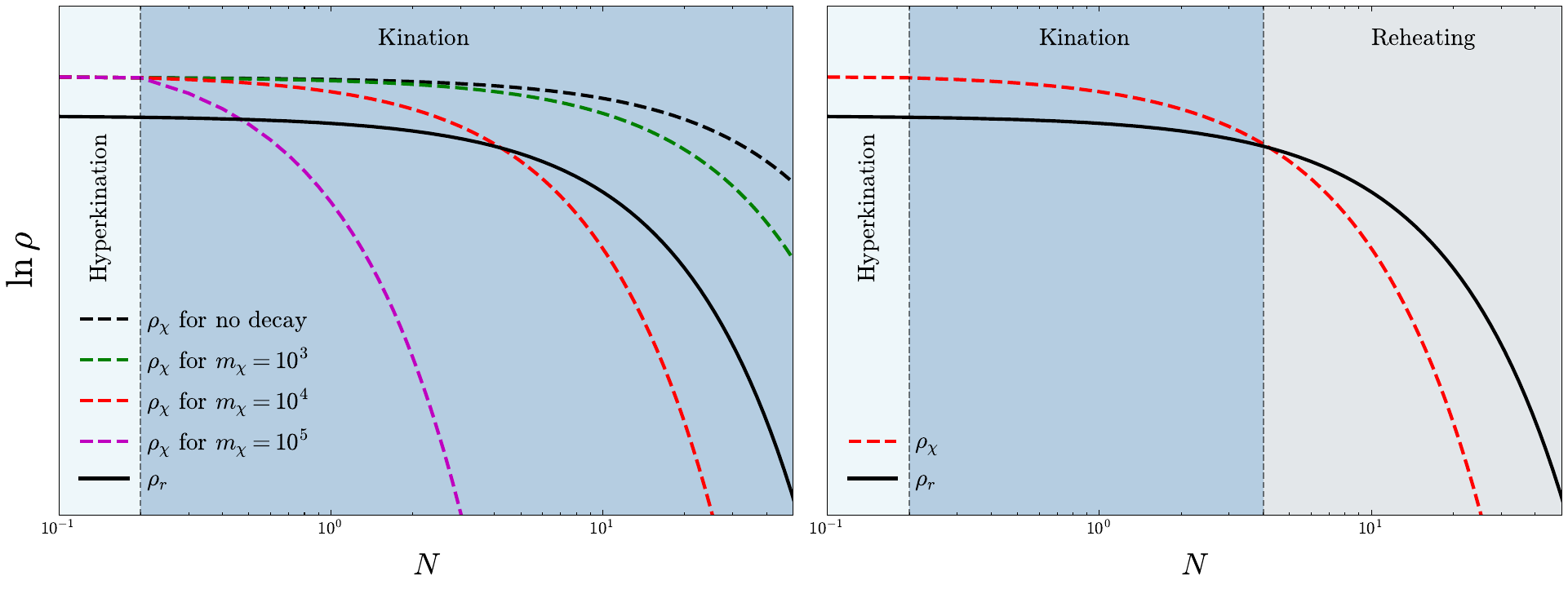}
    \caption{Left: Evolution of energy density of the scalar field and radiation with the number of $e$-folds for the case of no decay and considering a variety of cases for the mass of the scalar field (in GeV) for $t=2$. Right: The new modified evolution of energy density for radiation and the inflaton field. The length of hyperkination is fixed to the upper bound of $\Delta N_{\text{hyp}}=0.2$; for a choice of $m_\chi = 10^4$ GeV, the length of the kination phase is $\Delta N_{\text{kin}}=4$ for the coupling order $t=2$.}
    \label{fig:evol_reheating}
\end{figure}

To get an era of reheating, the mass of the inflaton field should be fine-tuned such that the decay rate causes the inflaton field to decay faster than radiation. This happens for $m_\chi = 10^4 - 10^5$ GeV, see the left panel of Figure \ref{fig:evol_reheating}.  If no decay process is considered (black dashed line), the scalar field energy density will never be sub-dominant than radiation. As we increase the mass parameter, the energy density dilutes faster, eventually letting radiation take over for $m_\chi > 8 \times 10^3$ GeV. Considering that the decay processes take time and do not happen as quickly as for $m_\chi = 10^5$ GeV, we choose the mass of the scalar field to be $10^4$ GeV. The length of hyperkination is fixed to the upper bound of $\Delta N_{\text{hyp}}=0.2$; for a choice of $m_\chi = 10^4$ GeV, the length of the kination phase is $\Delta N_{\text{kin}}=4$. The panel on the right of Figure \ref{fig:evol_reheating} is the new evolution of energy density accounting for the decay process in our alternate history theory for couplings with exponents $t>0$. The reheating temperature can be estimated by $T_{\text{reh}} \sim 0.1 \sqrt{\Gamma M_{\text{Pl}}} \sim 10^8$ GeV for $m_\chi = 10^4$ GeV giving a successful and efficient mechanism in such scenarios.

With the reheating mechanism in action, we assume a model where the cosmological evolution is as follows: inflation ($w=-1$), hyperkination ($w=1/3$), kination ($w=1$), and then reheating, after which the conventional eras of radiation ($w=1/3$) and matter domination ($w=0$) follow. The terms in the brackets denote the evolution of the barotropic parameter in this model. This non-standard expansion history brings forward new phenomenology, one of which is the alteration of the spectrum of primordial gravitational waves. We study those effects in the following sections.

\section{Gravitational Waves} \label{sec:GW}

Gravitational Waves (GWs) are a result of tensor perturbations. The primordial GWs originate as quantum vacuum fluctuations during inflation. The regularized gravitational wave energy density goes as follows,
\begin{equation} \label{eq:gw_energy_density}
    \langle \Hat{\rho}_{\text{GW}} \rangle \approx \int_{k=\mathcal{H}} \frac{(\mathrm{d} \ln k)}{\pi^2} \frac{k^4}{a^4} |\lambda_{-}|^2 
\end{equation}
where $k$ is the frequency of the wave vector, and $\mathcal{H} = a'/a$, where prime denotes derivative with respect to the conformal time $\eta$. Owing to the constraints on integration, we are confined to the sub-Hubble modes. In actuality, all modes of interest are significantly excited with $|\lambda_{-}| \gg 1$. In this limit, the vacuum contribution is insignificant, and the gravitational waves are fundamentally classical.

According to \eqref{eq:gw_energy_density}, it is evident that the energy density of gravitational waves at sub-Hubble scales behaves like radiation, with $\rho_{\text{GW}} \propto a^{-4}$, consistent with massless degrees of freedom. In cosmological theories characterized by a conventional expansion history, specifically those devoid of a hyperkination phase, only a small amount of gravitational waves are produced during inflation. These gravitational waves remain sub-dominant relative to the energy density of background radiation and the energy density of the field. During the phase of kinetic domination, the energy density of the field diminishes more rapidly than that of radiation, hence boosting the amount of gravitational wave energy density. The resultant gravitational wave spectrum is characterized by a peak and either conflicts with constraints on the number of relativistic degrees of freedom during Big Bang Nucleosynthesis or proves difficult to detect in gravitational wave experiments \cite{Tashiro_Chiba_Sasaki_2004,Sahni_Sami_Souradeep_2001,Riazuelo_Uzan_2000,Giovannini_1999,Figueroa_Tanin_2019,Artymowski_Czerwińska_Lalak_Lewicki_2018}. Having a period of hyperkination saves the theory from violating the BBN bound and thus demanding the need for such a phase and in turn $R^2$. The following section examines the modifications in the gravitational wave spectrum due to our theory.

\subsection{Background Solutions} \label{subsec:bg_soln}
To solve the background, we are particularly interested in the evolution of the scale factor $a$ as a function of the conformal time $\eta$. Throughout the cosmic history, the scale factor evolves during the following periods: inflation, hyperkination, kination, and radiation domination. Presumed to be instantaneous, the transition between these eras occur at conformal times $\eta^{\text{end}}_{\text{inf}} \equiv \eta_{\text{end}}$ (end of inflation and the start of hyperkination), $\eta_{\text{kin}}$ (start of kination and end of hyperkination), and $\eta_{\text{reh}}$ (start of reheating or radiation domination and end of kination). We employ identical indices to denote different variables assessed at these times. We demand the continuity of the scale factor and its first derivative (with respect to $\eta$) at those transition times.

We have,
\begin{equation} \label{eq:density_conformal}
    \mathrm{d} \eta = \frac{\mathrm{d} \tau}{a} = \frac{\mathrm{d} a}{a^2 H} = \frac{\mathrm{d} a}{a^2} \sqrt{\frac{3}{\rho}}
\end{equation}
Knowing the evolution of the energy density of the universe in $a$, we can invert \eqref{eq:density_conformal} and obtain $a(\eta)$ period by period. We normalize the scale factor, $a(\eta_{\text{end}}) = 1$, such that, for $a = e^N$, $N$ denotes the number of $e$-folds after the end of inflation. For inflation, we assume the generic slow roll condition, with $\eta_{\text{end}} < 0$ given by the condition,
\begin{equation}
    \epsilon \equiv -\frac{\Dot{H}}{H^2} = 1
\end{equation}
where $\epsilon$ is the first Hubble slow-roll parameter. Note that we will denote the Hubble parameter at the end of inflation $H_{\text{end}}$ as $H$. Solving for all the epochs, see Appendix \ref{app:scalefactor} for detailed derivation, we get
\begin{equation} \label{eq:scale_factor}
    a(\eta) =\begin{cases}
			\left[ \frac{-1}{(1-\epsilon)H \eta} \right]^{1/(1-\epsilon)} & \eta \leq \eta_{\text{end}}\\
            e^{\Delta N_{\text{hyp}}} \sin{ \left( \frac{H\eta + 1}{\sqrt{e^{2 \Delta N_{\text{hyp}} } - 1}} + \arcsin\left(e^{- \Delta N_{\text{hyp}}} \right) \right)} & \eta_{\text{end}} \leq \eta \leq \eta_{\text{kin}} \\
            \left( p \frac{H e^{-\Delta N_{\text{hyp}}}\sqrt{2c}}{\sqrt{e^{2\Delta N_{\text{hyp}}} -1 }} (\eta - \eta_{\text{kin}}) +  a_{\text{kin}}^{p} \right)^{1/p} & \eta_{\text{kin}} \leq \eta \leq \eta_{\text{reh}}\\
            H(\eta - \eta_{\text{reh}}) + a_{\text{reh}} & \eta_{\text{reh}} \leq \eta 
		 \end{cases}
\end{equation}
where,
\begin{align}
    \begin{split} \label{eq:p}
        p &= \frac{\Gamma \tau +4 - t\sqrt{6}}{2}
    \end{split}\\
    \begin{split} \label{eq:akin}
        a_{\text{kin}} & \equiv a(\eta_{\text{kin}}) = e^{\Delta N_{\text{hyp}}} \sin{ \left( \frac{H \eta_{\text{kin}} + 1}{\sqrt{e^{2 \Delta N_{\text{hyp}} } - 1}} + \arcsin\left(e^{- \Delta N_{\text{hyp}}} \right) \right)}
    \end{split}\\
    \begin{split} \label{eq:areh}
        a_{\text{reh}} & \equiv a(\eta_{\text{reh}}) = \left( p \frac{H e^{-\Delta N_{\text{hyp}}}\sqrt{2c}}{\sqrt{e^{2\Delta N_{\text{hyp}}} -1 }} (\eta_\text{reh} - \eta_{\text{kin}}) +  a_{\text{kin}}^{p} \right)^{1/p}
    \end{split}
\end{align}

We now want to estimate the conformal times at which the transitions occur. From \eqref{eq:transition_length}, we have,
\begin{equation} \label{eq:akin2}
    \Delta N_{\text{hyp}} - N_{\text{kin}} = \ln{\sqrt{2}} \implies e^{N_{\text{kin}}} = \frac{e^{\Delta N_{\text{hyp}}}}{\sqrt{2}} = a_{\text{kin}}
\end{equation}
Using \eqref{eq:akin} and \eqref{eq:akin2}, and solving for $\eta_{\text{kin}}$,
\begin{equation} \label{eq:etakin}
      \eta_{\text{kin}}  = \frac{1}{H} \left[ \sqrt{e^{2 \Delta N_{\text{hyp}} } - 1} \left( \frac{\pi}{4} - \arcsin\left(e^{- \Delta N_{\text{hyp}}} \right) \right) -1 \right]
\end{equation}

During kination, the total energy density goes as $\rho \propto a^{t\sqrt{6} - 6 - \Gamma \tau}$, whereas radiation scales as $\rho \propto a^{-4}$. The density parameter during kination thus goes as $\Omega_{r}^{\text{kin}} \propto a^{\Gamma \tau+2-t\sqrt{6}}$. During the era of reheating, the radiation dominates, i.e., $\Omega_{r}^{\text{reh}} \approx 1$. 
We can write,
\begin{equation}
    \Omega_{r}^{\text{reh}} \approx \Omega_{r}^{\text{kin}} \left( \frac{a_{\text{reh}}}{a_{\text{kin}}} \right)^{\Gamma \tau+2-t\sqrt{6}} \approx 1
\end{equation}

Consider, $\eta_{\text{reh}} \gg \eta_{\text{kin}}$, $\Omega_{r}^{\text{kin}} \approx \Omega_{r}^{\text{end}}$, and $a_{\text{kin}} = e^{\Delta N_{\text{hyp}}}/ \sqrt{2}$, we get,
\begin{equation}
    \Omega_{r}^{\text{end}} \approx \left( \frac{a_{\text{kin}}}{a_{\text{reh}}} \right)^{\Gamma \tau+2-t\sqrt{6}} \approx \left( \frac{a_{\text{kin}}}{a_{\text{reh}}} \right)^{2(p-1)} 
\end{equation}
Solving for $\eta_{\text{reh}}$ using \eqref{eq:areh} and \eqref{eq:akin2}, we get,
\begin{equation} \label{eq:etareh}
   \eta_\text{reh} \approx \frac{1}{H} \left( \left( \Omega_{r}^{\text{end}} \right)^{-\frac{p}{2(p-1)}} -1 \right) \left( \frac{e^{\Delta N_{\text{hyp}}}}{\sqrt{2}} \right)^{p+1}~ \frac{\sqrt{e^{2\Delta N_{\text{hyp}}} -1 }}{p}
\end{equation}
taking $c = 1$. We tune $\Gamma \tau$ such that we achieve reheating for $m_\chi = 10^4$ GeV and considered values of coupling constants. Thus, \eqref{eq:scale_factor} along with \eqref{eq:areh}, \eqref{eq:akin2}, \eqref{eq:etakin} and \eqref{eq:etareh} gives the evolution of the scale factor through time.

\subsection{Gravitational Wave Mode Functions} \label{subsec:gw_modefn}

The scale factor computed in the last section can be also written as,
\begin{equation}
    a(\eta) = \left( \frac{\eta}{\eta_c} \right)^{\frac{1}{2} - \nu} \hspace{0.3cm} , \hspace{0.3cm} \nu = \frac{3}{2}\left(1 - \frac{1}{2p} \right) w^2 - \frac{1}{2} \left( 1 + \frac{1}{p} \right) w + \frac{1}{4}\left( \frac{1}{p} - 2\right)
\end{equation}
where $\eta_c$ are corresponding constants from \eqref{eq:scale_factor}, $w$ is the corresponding barotropic parameter. During inflation, for de Sitter spacetime, $\nu = 3/2$ ($w =-1$); for more realistic quasi de Sitter spacetime, $\nu = 3/2 + \epsilon \equiv \nu_{\text{I}}$. During hyperkination, $\nu = -1/2$ ($w=1/3$), similar to that of the radiation-dominated era. While during kination, $\nu = 1/2 - 1/p \equiv \nu_{\text{II}}$ ($w=1$). We can write,
\begin{equation}
    \frac{a''}{a} = \left( \frac{1}{2} - \nu \right) \left( \frac{1}{2} - \nu -1 \right) \frac{1}{\eta^2} = \left( \nu^2 - \frac{1}{4} \right) \frac{1}{\eta^2}
\end{equation}
where prime denotes derivative with respect to the conformal time $\eta$. From the previous section, we can write,
\begin{equation}
    \frac{a''}{a} =\begin{cases}
			\frac{2+3\epsilon}{\eta^2} & \eta \leq \eta_{\text{end}}\\
            0 & \eta_{\text{end}} \leq \eta \leq \eta_{\text{kin}} \\
            \frac{1}{p} (\frac{1}{p} -1)\frac{1}{z^2} & \eta_{\text{kin}} \leq \eta \leq \eta_{\text{reh}}\\
            0 & \eta_{\text{reh}} \leq \eta 
		 \end{cases}
\end{equation}
where,
\begin{equation}
    z \equiv \eta - \frac{\eta_{\text{kin}}}{2} + \frac{1}{H}
\end{equation}
We see that $a''=0$ during the hyperkination and radiation-dominated era. This shows that the spectrum from kination is truncated by upper and lower limits. 

For mode function, $v_k^{s}$, with polarizations $s = \oplus, \otimes$ and frequency $k$, we can write the Mukhanov-Sasaki equation as,
\begin{equation}
    v^{s''}_k + \left( k^2 - \frac{a''}{a} \right)v^s_k = 0 \implies v^{s''}_k + \left[ k^2 - \left( \nu^2 - \frac{1}{4} \right) \frac{1}{\eta^2} \right] v^s_k = 0
\end{equation}
We perform a change of variables $x = k\eta$ (when $\eta<0$, $x = -k\eta$ during inflation). We drop $s$ and $k$ to avoid clutter.
\begin{equation}
    \frac{\mathrm{d}^2 v}{\mathrm{d}x^2} + \left[ 1 - \left( \nu^2 - \frac{1}{4} \right)\frac{1}{x^2} \right] v = 0
\end{equation}
We now redefine the mode functions $v = \sqrt{x}g$, which recasts the equation as a Bessel equation.
\begin{equation}
    0 = x^2 \frac{\mathrm{d}^2 g}{\mathrm{d}x^2} + x \frac{\mathrm{d}g}{\mathrm{d}x} + g \left( x^2 - \nu^2 \right)
\end{equation}
The most general solution to the Bessel equation is given by,
\begin{align}
    \begin{split}
        g(x) &= c_1 H_\nu^{(1)}(x) + c_2 H_\nu^{(2)}(x)
    \end{split} \\
    \begin{split} \label{eq:MS_mostgensol}
        v(x) &= \sqrt{x} \left( c_1 H_\nu^{(1)}(x) + c_2 H_\nu^{(2)}(x) \right)
    \end{split}
\end{align}
where $H_\nu^{(1)}$ and $H_\nu^{(2)}$ are Hankel functions of the first and second kind, respectively. We solve for the mode functions in detail in Appendix \ref{app:modefn} and note the final results here,
\begin{equation} \label{eq:MS_solutions}
    v^s_k(\eta) = \begin{cases}
			\sqrt{\frac{\pi}{4k}} \sqrt{-k \eta} ~e^{\frac{i\pi}{4}(1+2 \nu_I)} H_{\nu_I}^{(1)}(-k \eta) & \eta \leq \eta_{\text{end}}\\
            \frac{1}{\sqrt{2k}} \left[ \alpha_+(k) e^{-ik\eta} + \alpha_-(k) e^{ik\eta}\right] & \eta_{\text{end}} \leq \eta \leq \eta_{\text{kin}} \\
            \sqrt{\frac{\pi z}{4}} \left[ \beta_+(k) ~e^{-\frac{i\pi}{4}(1+2 \nu_{II})} H_{\nu_{II}}^{(2)}(kz) + \beta_-(k) ~e^{\frac{i\pi}{4}(1+2 \nu_{II})} H_{\nu_{II}}^{(1)}(kz)\right] & \eta_{\text{kin}} \leq \eta \leq \eta_{\text{reh}}\\
            \frac{1}{\sqrt{2k}} \left[ \gamma_+(k) e^{-ik\eta} + \gamma_-(k) e^{ik\eta}\right] & \eta_{\text{reh}} \leq \eta 
		 \end{cases}
\end{equation}

To determine the constants $\alpha_\pm$, $\beta_\pm$, and $\gamma_\pm$; we demand the continuity of $v_k^s$ and its derivative at the transition times $\eta_{\text{end}}$, $\eta_{\text{kin}}$, $\eta_{\text{reh}}$. The details for this mode-matching process are mentioned in Appendix \ref{app:modematching}. The results are as follows,
\begin{align} 
    \begin{split} \label{eq:MS_coeff_alpha}
        \alpha_\pm &= \mp \frac{f(\epsilon)}{2} \left( \frac{H}{k} \right)^{2 + \epsilon}
    \end{split}\\
    \begin{split} \label{eq:MS_coeff_beta}
        \beta_\pm &= i e^{\pm \frac{i\pi}{4}(1+2 \nu_{II})}~\alpha_-  \beta_0 (k\eta_{\text{kin}})^{1/p}
    \end{split}\\
    \begin{split} \label{eq:MS_coeff_gamma}
         \gamma_\pm &= \mp 4 \alpha_- \left( \frac{\eta_{\text{kin}}}{\eta_{\text{reh}}}\right)^{1/p}
    \end{split}
\end{align}
where, 
\begin{equation}
    f(\epsilon) \equiv 2^{\epsilon} e^{\frac{i\pi}{2}\epsilon}   (\epsilon + 1) \frac{\Gamma ( 3/2 + \epsilon )}{\Gamma ( 3/2 )} \quad \text{and} \quad \beta_0 \equiv  2^{2\nu_{II}+1} \frac{\Gamma(3/2-1/p)}{\Gamma(3/2)}
\end{equation}

In pure de Sitter case, $\epsilon \to 0$, which gives $f(\epsilon) \to 1$, then the moduli squared of those coefficients take the form,
\begin{align}
    |\alpha_\pm|^2 & = \frac{H^4}{4k^4}\\
    |\beta_\pm|^2 & = \beta_0^2 \frac{H^4}{4k^4} (k ~\eta_{\text{kin}})^{2/p}\\
    |\gamma_\pm|^2 & = \frac{H^4}{k^4}  \left( \frac{\eta_{\text{kin}}}{2\eta_{\text{reh}}}\right)^{2/p}
\end{align}

We now move to deduce the metric perturbations, $h^s_k$, from the solutions of the Mukhanov-Sasaki equation in \eqref{eq:MS_solutions} and \eqref{eq:MS_coeff_alpha}-\eqref{eq:MS_coeff_gamma}. The metric perturbations are given as,
\begin{equation}
    h^s_k (\eta) = \frac{\sqrt{2} v^s_k}{a M_{\text{Pl}}}
\end{equation}
where $a$ is the scale factor corresponding to the respective epoch. Solving for $h^s_k (\eta)$, we get,
\begin{equation}
    h^s_k(\eta) = \begin{cases}
            \frac{i}{ M_{\text{Pl}} \sqrt{k}} \frac{\sin{(k \eta)}}{a(\eta)} \frac{f(\epsilon)}{2} \left( \frac{H}{k} \right)^{2 + \epsilon}& \eta_{\text{end}} \leq \eta \leq \eta_{\text{kin}} \\
             \frac{i f(\epsilon)}{M_{\text{Pl}} a(\eta)} \left( \frac{H}{k} \right)^{2+\epsilon} \frac{\Gamma(3/2-1/p)}{\Gamma(3/2)} \frac{4}{\sqrt{k}} \left( \frac{k\eta_{\text{kin}}}{2} \right) ^{1/p} \cos{\left( k\eta + \frac{\pi}{2} \left( \frac{1}{p} - 1 \right) \right)}& \eta_{\text{kin}} \leq \eta \leq \eta_{\text{reh}}\\
            \frac{4i}{M_{\text{Pl}} \sqrt{k}} \frac{\sin{(k \eta)}}{a(\eta)} \frac{f(\epsilon)}{2} \left( \frac{H}{k} \right)^{2 + \epsilon} \left( \frac{\eta_{\text{kin}}}{\eta_{\text{reh}}}\right)^{1/p} & \eta_{\text{reh}} \leq \eta 
		 \end{cases}
\end{equation}

\begin{figure}
    \centering
    \includegraphics[width=\textwidth]{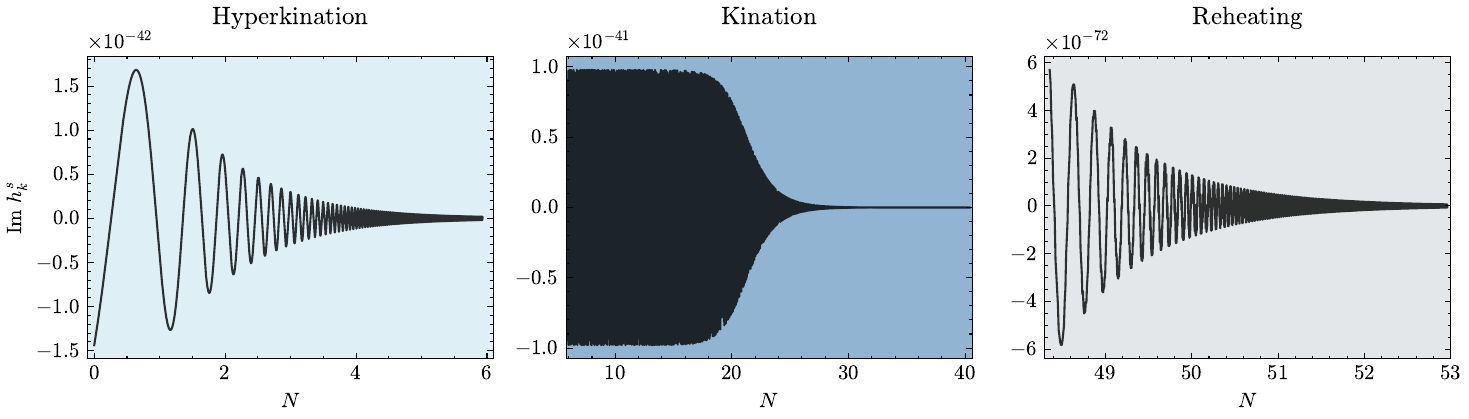}
    \caption{The imaginary part of the perturbations are plotted against the number of $e$-folds passed after inflation $N_{\text{end}}=0$ for the mode frequency $k=10^{-5}\times M_{\text{Pl}}$. For this analysis, we use $\Delta N_{\text{hyp}}=6.18$ from Table \ref{tab:no_of_efolds} for the coupling $t=2$. During hyperkination (left), the gravitational waves produced quickly die down. However, as soon as the kination phase starts (center), the model acquires a large amount of energy, and the amplitude of perturbations increases. These modes then slowly lose energy as the kination phase ends and then continue to decay during and beyond the reheating as well (right). We used $\Omega_r^{\text{end}} = 10^{-10}$ and $H=10^{13}$ GeV.}
    \label{fig:tensor-gw}
\end{figure}

We depict the analytical solutions of these metric perturbations in Figure \ref{fig:tensor-gw}. The imaginary part of the perturbations is plotted against the number of $e$-folds passed after inflation $N_{\text{end}}=0$ for the phases of hyperkination (left panel), kination (middle panel) and reheating and beyond (right panel). It is apparent that during the phase of hyperkination, the gravitational waves produced quickly die down. However, as soon as the kination phase starts, the model acquires a large amount of energy, and the amplitude of perturbations increases by an order of one. These modes then slowly lose a large amount of energy as the kination phase ends and then continue to decay beyond reheating as well.

\begin{figure}
    \centering
    \includegraphics[width=\textwidth]{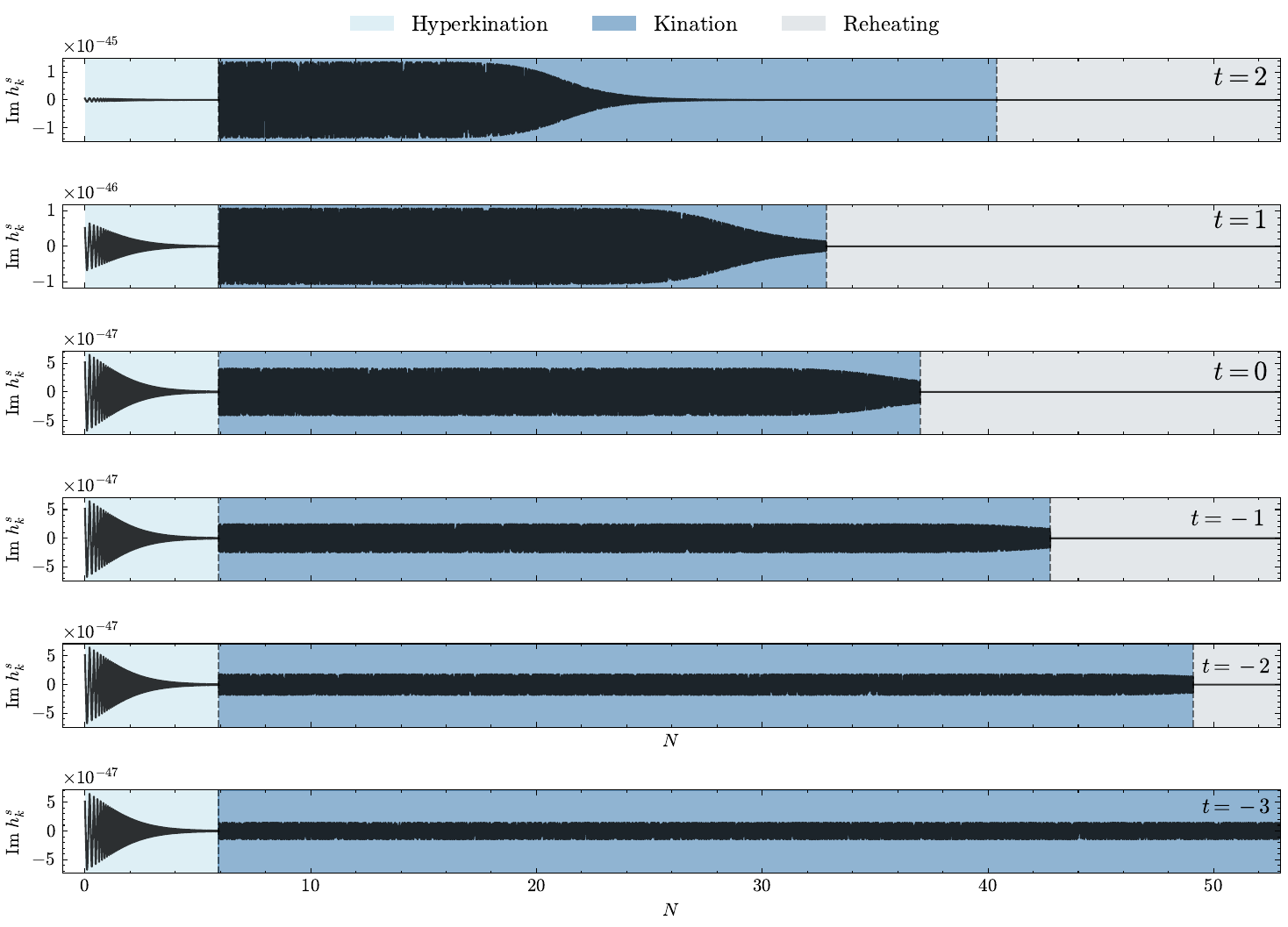}
    \caption{The imaginary part of the perturbations are plotted against the number of $e$-folds passed after inflation $N_{\text{end}}=0$ for the mode frequency $k=10^{-4}\times M_{\text{Pl}}$ for different orders of coupling, keeping $\Delta N_{\text{hyp}}=6.18$ constant. For kination, we see that the perturbations get a boost in energy for all orders. However, as the order decreases, the amplitude of produced gravitational wave modes increases, indicating that the hyperkination phase will contribute significantly to the primordial background. We used $\Omega_r^{\text{end}} = 10^{-10}$ and $H=10^{13}$ GeV for all panels.}
    \label{fig:tensor_gw_coupling}
\end{figure}

\begin{figure}
    \centering
    \includegraphics[width=\textwidth]{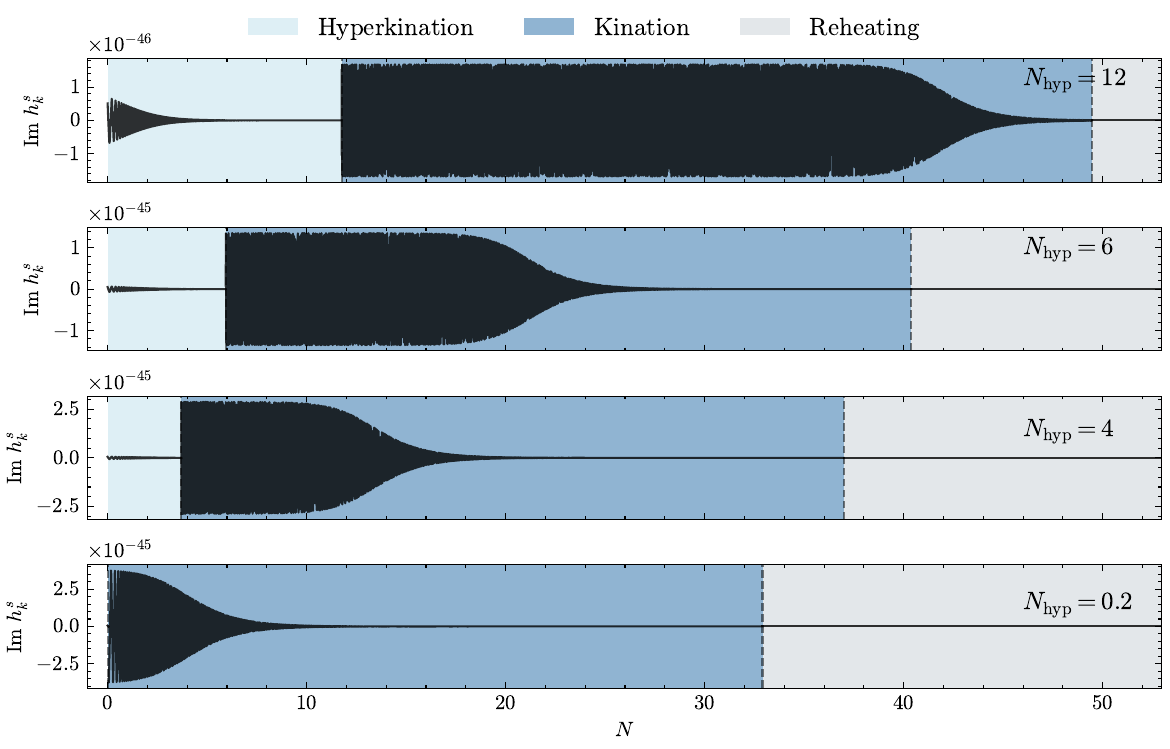}
    \caption{The imaginary part of the perturbations are plotted against the number of $e$-folds passed after inflation $N_{\text{end}}=0$ for the mode frequency $k=10^{-4}\times M_{\text{Pl}}$ for different lengths of hyperkination, keeping the order of coupling $t=2$ constant.  We do not observe any change in GW perturbations generated during hyperkination. However, the kination phase reduces in length, and the modes decay faster with decreasing length of hyperkination. We used $\Omega_r^{\text{end}} = 10^{-10}$ and $H=10^{13}$ GeV for all panels.}
    \label{fig:tensor_gw_nhyp}
\end{figure}

We check the time-evolution of the perturbations by fixing the length of the period of hyperkination and varying the order of coupling; see Figure \ref{fig:tensor_gw_coupling}. We keep $\Delta N_{\text{hyp}} = 6.18$ so that we can gauge the changes if there are any from the variation. The order of the coupling decreases as we go from top to bottom. Similar to the results obtained in Figure \ref{fig:tensor-gw} for kination, we see that the perturbations get a boost in energy for all orders. However, as the order decreases, the amplitude of produced gravitational wave modes during hyperkination increases, indicating that the hyperkination phase will contribute significantly to the primordial background for such couplings. Another feature is the increasing length of the period of kination with decreasing order. This suggests that the energy from GWs during kination will also increase for every decreasing order. In Figure \ref{fig:tensor_gw_nhyp}, we vary $\Delta N_{\text{hyp}}$ for the $t=2$ coupling. We do not observe any change in GW perturbations generated during hyperkination. However, the kination phase reduces in length, and the modes decay faster with decreasing length of hyperkination.

From these observations, we can draw a few conclusions: 1. Coupling of lesser order boosts the perturbations during hyperkination. 2. The kinationary phase gives a boost to primordial GWs for all orders. 3. Having a smaller duration of hyperkination dilutes the contribution of GWs from kination. While we focused on one particular mode $k$, other modes will also contribute to the energy density from these primordial GWs. We probe to find this spectral energy density in the following section.

\subsection{Gravitational Wave Spectrum} \label{subsec:gw_spectrum}

We now move towards calculating the spectral energy density of the primordial gravitational waves (GWs) background. It is defined as,
\begin{equation}
    \Omega_{\text{GW}} (k, \eta_0) \equiv \frac{1}{\rho (\eta)} \frac{\mathrm{d} \rho_{\text{GW}}(k, \eta)}{\mathrm{d} \ln k} = \frac{1}{\rho (\eta)} \frac{k^4 |\lambda_- (k)|^2}{\pi^2 a^4(\eta)}
\end{equation}
where $\rho$ represents the total energy density of the Universe and $\rho_{\text{GW}}(k, \eta)$ denotes the contribution to the gravitational wave energy density from modes around $k$, as specified by \eqref{eq:gw_energy_density} for the predominant, sub-Hubble modes. We match $\lambda_-$ with $\alpha_-$, $\beta_-$, or $\gamma_-$. The radiation energy density can be expressed with the normalization $a_\text{end} = 1$
\begin{equation} \label{eq:radiation_ed}
    \rho_r (\eta) = \Omega_r (\eta) \rho (\eta) =  \Omega_r^\text{end} \rho_\text{end} a^{-4}(\eta) \implies \rho (\eta) ~a^{4} (\eta) = \rho_\text{end} \frac{ \Omega_r^\text{end}}{\Omega_r (\eta)}
\end{equation}
We want to calculate the spectral energy density today. Computing the spectrum today\footnote{We use 0 to index quantities today.}, $\eta = \eta_0$,
\begin{equation}
    \Omega_{\text{GW}} (k, \eta_0) = \frac{1}{\rho_\text{end}} \frac{ \Omega_r^0}{\Omega_r^\text{end}} \frac{k^4}{\pi^2}  |\lambda_- (k)|^2
\end{equation}

\begin{equation}
    \Omega_{\text{GW}} (k, \eta_0) = \begin{cases}
            \frac{ \Omega_r^0}{\Omega_r^\text{end}} \frac{H^2}{M_{\text{Pl}}^2} \frac{1}{3\pi^2}  \left( \frac{\eta_{\text{kin}}}{2\eta_{\text{reh}}}\right)^{2/p} & k < k_{\text{reh}} \\
           \frac{ \Omega_r^0}{\Omega_r^\text{end}} \frac{H^2}{M_{\text{Pl}}^2} \frac{1}{12 \pi^2}  \beta_0^2 (k ~\eta_{\text{kin}})^{2/p} & k_{\text{reh}} < k < k_{\text{kin}}\\
           \frac{ \Omega_r^0}{\Omega_r^\text{end}} \frac{H^2}{M_{\text{Pl}}^2} \frac{1}{12 \pi^2}  & k_{\text{kin}} < k < k_{\text{end}}
		 \end{cases}
\end{equation}
where, $\rho_{\text{end}}= 3H^2 M_{\text{Pl}}^2$. We get the boundary values for $k$ at the end of inflation and transition times using the condition $\eta_{\text{reh}} \gg \eta_{\text{kin}} \gg |\eta_{\text{end}}|$,
\begin{equation}
    k_{\text{end}} = \frac{1}{\eta_{\text{end}}} = H, \quad
    k_{\text{kin}} = \frac{2}{\eta_{\text{kin}}}, \quad
    k_{\text{reh}} = \frac{1}{\eta_{\text{reh}}} 
\end{equation}

We compute the spectrum as a function of $f$, the GW frequency today. To relate $f$ to wavenumber $k$, we use \eqref{eq:radiation_ed} and $\rho = 3H^2 M_{\text{Pl}}^2$, which gives
\begin{equation}
    f = \frac{k}{2 \pi a_0} = \frac{1}{2 \pi} \left( \frac{\Omega_r^0}{\Omega_r^{\text{end}}} \frac{H_0^2}{H^2} \right)^{1/4} k
\end{equation}

An important frequency is that which relates to Big Bang Nucleosynthesis (BBN). It is independent of the early expansion history, and we can solve it directly as
\begin{equation}
    f_{\text{BBN}} = \frac{1}{2 \pi} \frac{a_{\text{BBN}}}{a_0} H_{\text{BBN}}= \frac{1}{2 \pi} \left( \frac{\rho_r^0}{\rho_{\text{BBN}}} \right)^{1/4} \left( \frac{\rho_{\text{BBN}}}{3 M_{\text{Pl}}^2}\right)^{1/2} \approx 1.36 \times 10^{-11} \text{Hz}
\end{equation}
where we used $\rho_{\text{BBN}} \approx 3 \times 10^{-86}$ $M_{\text{Pl}}^4$ \cite{Sanchez_Lopez_2023}. We can draw a relation between our original free parameter $\alpha$, energy density at the end of inflation $\rho_{\text{end}}$ and $\Delta N_{\text{hyp}}$ using \eqref{eq:rho-chi} and \eqref{eq:deltaN}
\begin{align}
    \alpha \rho_\text{end} &= \frac{6 - \chi_0^{2'} e^{2N}}{\kappa h \chi_0^{4'} e^{4N}} = \frac{1 - e^{-2\Delta N_{\text{hyp}}}}{6 \kappa h~ e^{-4\Delta N_{\text{hyp}}}} = 3 \alpha H^2\\
    \begin{split} \label{eq:N_alpha}
        e^{\Delta N_{\text{hyp}}} &= \left( \frac{1 + \sqrt{1 + 72 \alpha \kappa h H^2}}{2} \right)^{1/2}
    \end{split}
\end{align}

Using all the relations, we get the GW spectrum in terms of frequency $f$ computed today as,
\begin{equation} \label{eq:gwspectrum}
    \Omega_{\text{GW}} (f, \eta_0) = \begin{cases}
            \frac{ \Omega_r^0}{\Omega_r^\text{end}} \frac{H^2}{M_{\text{Pl}}^2} \frac{1}{3\pi^2}  \left( \frac{\eta_{\text{kin}}}{2\eta_{\text{reh}}}\right)^{2/p} & f < f_{\text{reh}} \\
           \frac{ \Omega_r^0}{\Omega_r^\text{end}} \frac{H^2}{M_{\text{Pl}}^2} \frac{1}{12 \pi^2}  \beta_0^2 (2 \pi f~\eta_{\text{kin}})^{2/p} \left( \frac{\Omega_r^{\text{end}}}{\Omega_r^0} \frac{H^2}{H_0^2} \right)^{1/2p}& f_{\text{reh}} < f < f_{\text{kin}}\\
           \frac{ \Omega_r^0}{\Omega_r^\text{end}} \frac{H^2}{M_{\text{Pl}}^2} \frac{1}{12 \pi^2}  & f_{\text{kin}} < f < f_{\text{end}}
		 \end{cases}
\end{equation}
where 
\begin{align}
    f_{\text{reh}} &= \frac{1}{2 \pi} \left( \frac{\Omega_r^0}{\Omega_r^{\text{end}}} \frac{H_0^2}{H^2} \right)^{1/4} k_{\text{reh}} = \frac{1}{2 \pi} \left( \frac{\Omega_r^0}{\Omega_r^{\text{end}}} \frac{H_0^2}{H^2} \right)^{1/4} \frac{1}{\eta_{\text{reh}} }\\
    f_{\text{kin}} &= \frac{1}{2 \pi} \left( \frac{\Omega_r^0}{\Omega_r^{\text{end}}} \frac{H_0^2}{H^2} \right)^{1/4} k_{\text{kin}} = \frac{1}{\pi} \left( \frac{\Omega_r^0}{\Omega_r^{\text{end}}} H_0^2 H^2 \right)^{1/4} \frac{1}{\eta_{\text{kin}} } \\
    f_{\text{end}} &= \frac{1}{2 \pi} \left( \frac{\Omega_r^0}{\Omega_r^{\text{end}}} \frac{H_0^2}{H^2} \right)^{1/4} k_{\text{end}} = \frac{1}{2 \pi} \left( \frac{\Omega_r^0}{\Omega_r^{\text{end}}} H_0^2 H^2 \right)^{1/4}
\end{align}
with $p$, $\eta_{\text{kin}}$ and $\eta_{\text{reh}}$ given by \eqref{eq:p}, \eqref{eq:etakin} and \eqref{eq:etareh} respectively.

The spectrum features two plateaus, one during hyperkination and one after reheating. These plateaus \textit{sandwich} the epoch of kination, which gives a \textit{boost} to the primordial gravitational waves. The enhancement is observed as $\Omega_{\text{GW}} \propto f^{2/p}$, where the value of $p$ changes as we change the order of coupling $t$.

\section{Gravitational Wave Observations} \label{sec:GWObs}

We check for the observability of gravitational waves from the epochs of hyperkination, kination, and reheating for a number of variable parameters such as $t$, $\alpha$, $\Omega_r^{\text{end}}$ and $H$. We compare the spectral energy density today $\Omega_{\text{GW}}(f,\eta_0)$ against the power-law integrated sensitivity curves (PLICs) for a variety of current and future, ground and space-based GW detectors.

PLICs are detector sensitivity curves for stochastic gravitational-wave backgrounds that take into account the increase in sensitivity that comes from integrating over frequency in addition to integrating over time \cite{Thrane_2013}. The plots in this section show the PLIC for different detectors as shaded backgrounds and the spectrum from primordial gravitational waves as solid lines. The plots can be interpreted as follows: If a curve for a predicted GW background lies \textit{below} the PLIC, then for such a background, the sound-to-noise ratio (SNR) for that particular detector is $<1$. While, if it lies anywhere \textit{above} the PLIC, then the background will be observed in the detector with SNR $>1$. PLIC curves for LIGO (O3) and LIGO (O5) are generated using the formalism mentioned in \cite{Thrane_2013}; curves BBO, CE, ET, DECIGO, LISA, PTAs, SKA, and NANOGrav is obtained from \cite{Schmitz_2021} while the one for Resonant Cavities is from \cite{Herman_2023}. The PLICs for ASTROD-GW and $\mu$Ares were obtained from their respective noise spectra \cite{ASTRODGW_2013,muAres}.

\subsection{Minimal $R^2$ with $\Delta N_{\text{hyp}}=0.2$}
\begin{figure}[t]
    \centering
    \includegraphics[width=\textwidth]{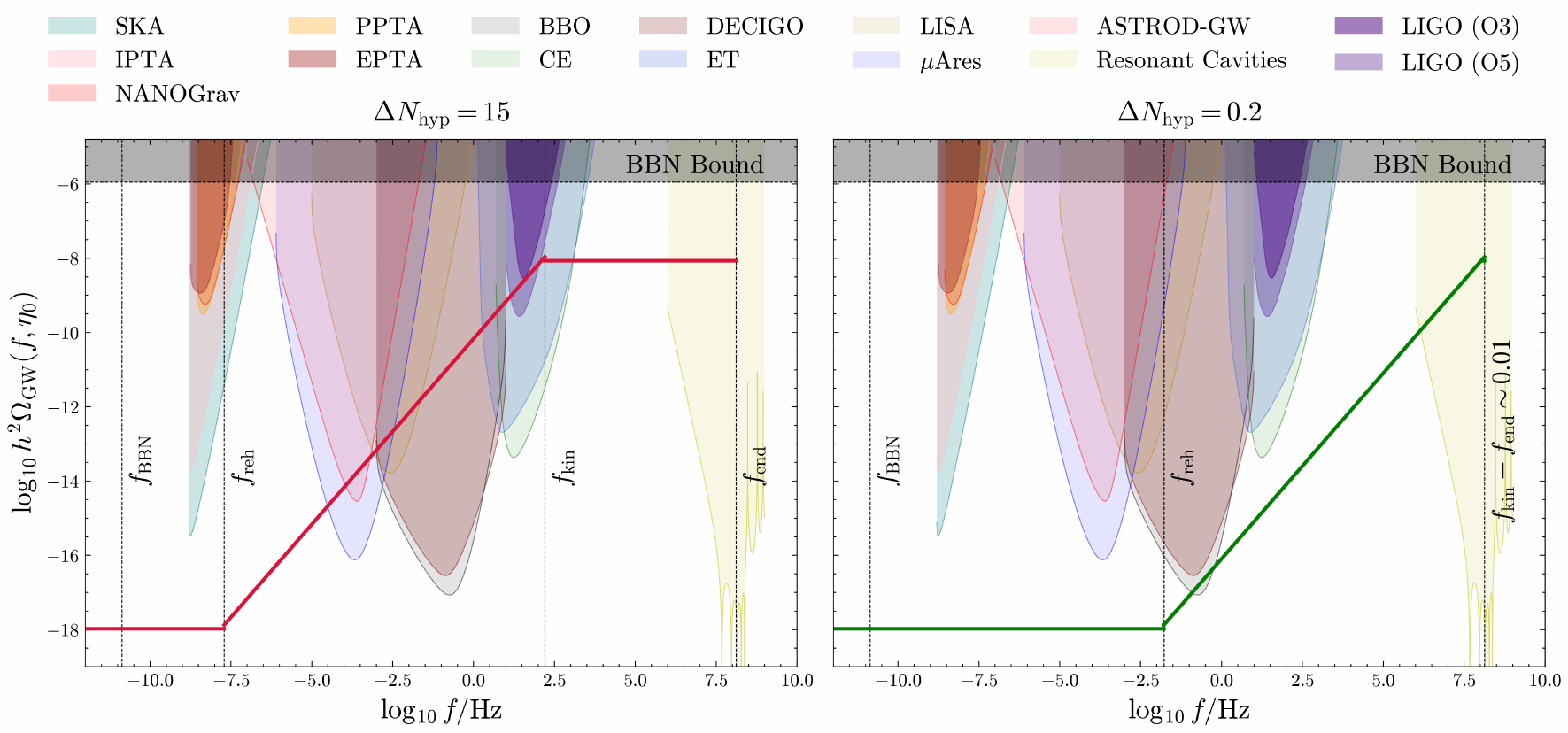}
    \caption{The left panel shows the observability considering $\Delta N_{\text{hyp}}=15$, while the right panel takes $\Delta N_{\text{hyp}}=0.2$. A shorter period of hyperkination gives a longer period of kination and reheating. While the order of magnitude of the boost is the same, the spectrum gets enhanced in a different frequency range, making it less detectable by future GW experiments. The gray region above highlights the BBN bound, while the vertical dashed lines show the frequency of BBN, reheating, kination, and end of inflation or the start of hyperkination. We consider $H = 10^{13}$ GeV and $\Omega_r^{\text{end}} = 10^{-10}$.}
    \label{fig:gwobs_Nhyp}
\end{figure}

A study considered a model with 
$\mathcal{L} \propto h(\chi)R + \alpha R^2$, where $h(\chi) = 1 + \chi^2$ \cite{Sanchez_Lopez_2023}. They claimed that having a longer period of hyperkination boosts the GW spectrum to an observable regime. However, from the energy density argument, the length of hyperkination is bounded from above, and thus, considering a longer period will destabilize the universe. We probe the observability of GWs from their model but with $\Delta N_{\text{hyp}}=0.2$. The left panel in the Figure \ref{fig:gwobs_Nhyp} shows the observability considering $\Delta N_{\text{hyp}}=15$, while the right panel takes $\Delta N_{\text{hyp}}=0.2$. A difference is apparent from the figure, a shorter period of hyperkination gives longer kination and longer period for reheating. While the order of magnitude of the boost is the same, the spectrum gets enhanced in a different frequency range, making it less detectable by future GW experiments.

\begin{figure}[h]
    \centering
    \includegraphics[width=\textwidth]{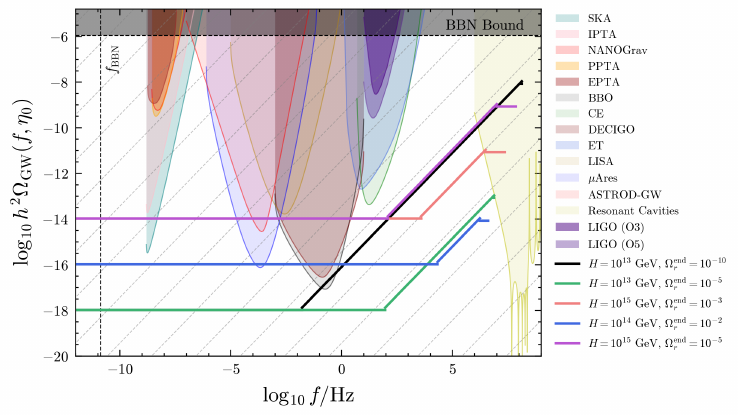}
    \caption{Spectrum of primordial GW for $\mathcal{L} \propto h(\chi)R + \alpha R^2$, where $h(\chi) = 1 + \chi^2$, with the change $\Delta N_{\text{hyp}}=0.2$ for a variety of values for the Hubble parameter, $H$, and the energy density of radiation, $\Omega_r^{\text{end}}$, at the end of inflation. The spectrum is plateaued after reheating and grows as $\Omega_{\text{GW}} \propto f$ during kination. The gray region above highlights the BBN bound, while the vertical dashed line shows the frequency of BBN. The diagonal dashed gray lines show the boost for different values of $H$ and $\Omega_r^{\text{end}}$.}
    \label{fig:gwobs_H}
\end{figure}

Considering $\Delta N_{\text{hyp}}=0.2$ for their model, we probe the detectability for a few different parameter values for $H$ and $\Omega_r^{\text{end}}$ in Figure \ref{fig:gwobs_H}. The spectrum is plateaued after reheating and grows as $\Omega_{\text{GW}} \propto f$ during kination. The peak gets truncated owing to the era of hyperkination, which prevents it from violating the BBN bound. While a small part of the spectrum falls in the detectable region of GW experiments, any observation of such kind from the stochastic background will hint towards the presence of $R^2$ terms.

\subsection{Available Parameter Space}

\begin{figure}[t]
    \centering
    \includegraphics[width=\textwidth]{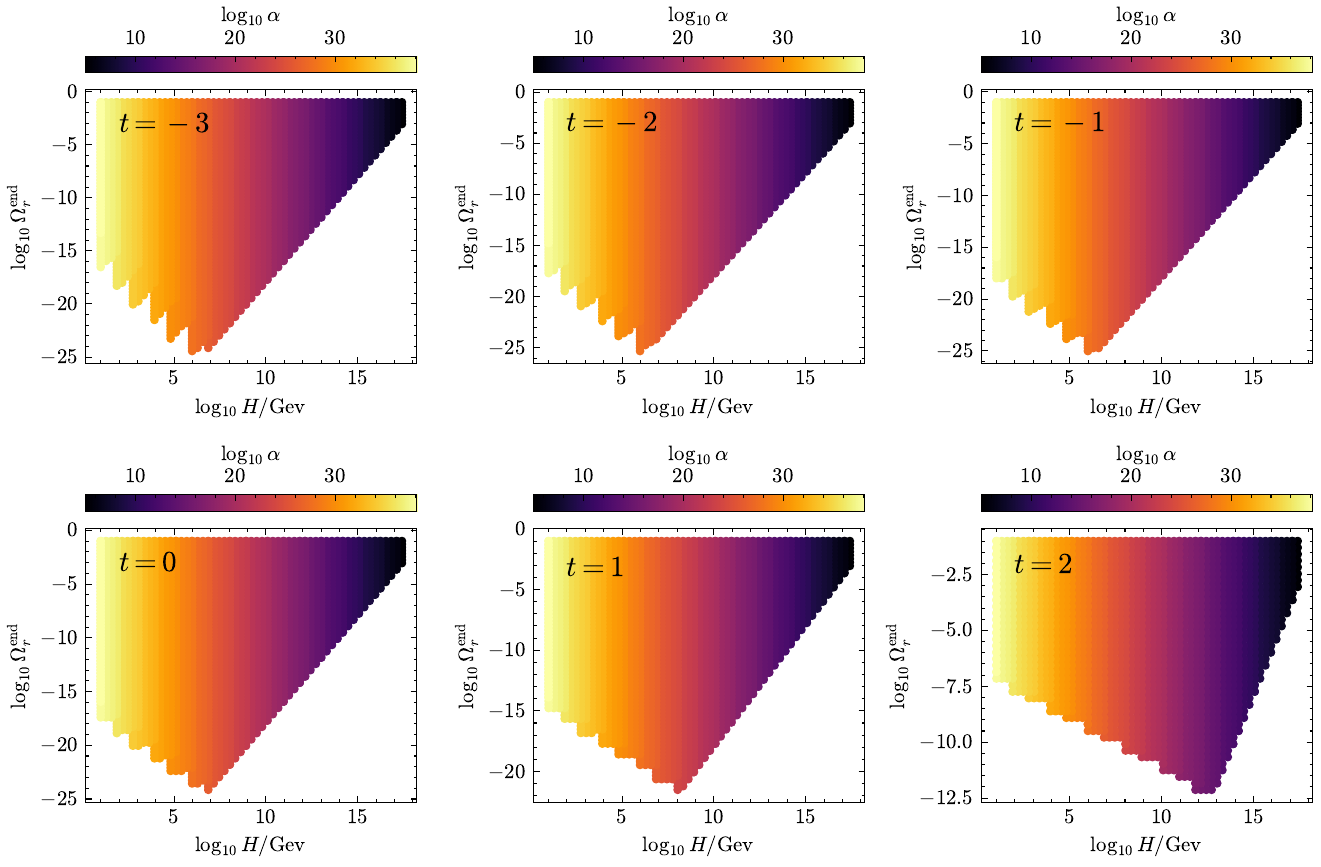}
    \caption{Available parameter space for $\alpha$, $H$ and $\Omega_{r}^{\text{end}}$ for different orders of coupling.}
    \label{fig:paramspace}
\end{figure}

We transferred the dependence of the spectrum from the length of hyperkination $\Delta N_{\text{hyp}}$ to $\alpha$  and $H$, which was the initial free parameter using \eqref{eq:N_alpha}.  We are also free to vary $H$ and $\Omega_r^{\text{end}}$, which are the Hubble parameter and energy density of radiation at the end of inflation, respectively. Considering the evolution of the universe to be sequential, we had assumed that,
\begin{equation} \label{eq:constraint1}
    \eta_{\text{end}} \ll \eta_{\text{hyp}} \ll \eta_{\text{kin}} \ll \eta_{\text{reh}} 
\end{equation}
Also,
\begin{equation} \label{eq:constraint2}
    f_{\text{BBN}} < f_{\text{reh}} < f_{\text{kin}} < f_{\text{end}} 
\end{equation}
The hyperkination phase acts as an upper bound for the boost to the GW spectrum, and thus it requires that,
\begin{equation} \label{eq:constraint3}
    \Omega_{\text{GW}}^{\text{hyp}} < \Omega_{\text{GW}}^{\text{BBN}}
\end{equation}
This requirement further constrains the permissible values for $H$ and $\Omega_{r}^{\text{end}}$. These constraints, \eqref{eq:constraint1}, \eqref{eq:constraint2}, and \eqref{eq:constraint3}, must be followed by any theory for it to be valid. From Table \ref{tab:no_of_efolds}, it is apparent that the point of convergence varies for different theories, and hence the length of hyperkination changes. As $\Delta N_{\text{hyp}} \leq 0.2$, it does not agree with the obtained convergence. The role of the hyperkinationary phase is to not let the boost during kination violate the BBN bound and, thus, having a slightly longer period of hyperkination since all the values of convergence are small compared to the required number of $e$-folds for inflation will not affect the spectrum much. For any value of $H$ and $\Omega_{r}^{\text{end}}$ we consider a small value of $\alpha$ (and thus $\Delta N_{\text{hyp}}$) from the available parameter space to bypass this issue. Taking into account all the constraints mentioned above, we probe the parameter space in Figure \ref{fig:paramspace} for available values of $\alpha$, $H$ and $\Omega_{r}^{\text{end}}$ for different orders of couplings such that the theory stays valid.

\subsection{Non-minimal $R^2$}
\begin{figure}[]
    \centering
    \includegraphics[width=\textwidth]{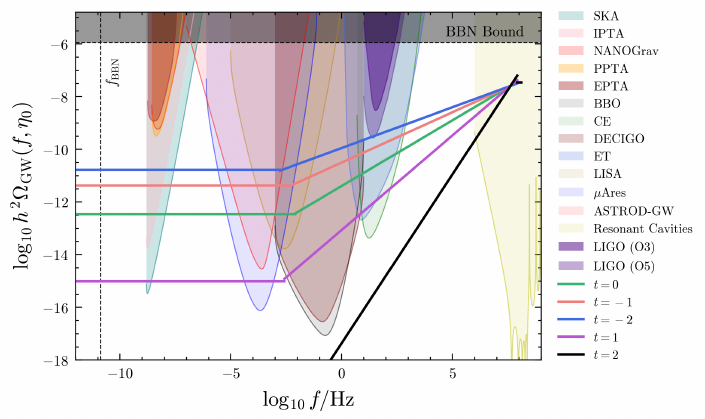}
    \caption{Spectrum of primordial GW for $\mathcal{L} \propto h(\chi)(R + \alpha R^2)$, where $h(\chi) = (1 + \xi \chi /M_{\text{Pl}})^t$, for a variety of coupling orders $t$. A boost in the GW spectrum during kination, $\Omega_{\text{GW}} \propto f^{2/p}$, is observed with a varying slope for all considered couplings. The spectrum after reheating and during hyperkination is flat. The hyperkinationary era prevents the enhancement in the spectrum from violating the BBN bound. The spectrum falls more into an observable regime with a decreasing order of coupling. The gray region above highlights the BBN bound, while the vertical dashed line shows the frequency of BBN. We considered $\alpha=10^{14}$, $H = 10^{13}$ GeV and $\Omega_r^{\text{end}} = 10^{-10}$.}
    \label{fig:gwobs_t}
\end{figure}

Our model considers a Lagrangian with $\mathcal{L} \propto h(\chi)(R+\alpha R^2)$, where $h(\chi)=(1+ \xi \chi /M_{\text{Pl}})^t$ with $t \leq 2$ being the order of coupling. We obtained the spectrum as mentioned in \eqref{eq:gwspectrum}. For our case, the enhancement during kination is observed as $\Omega_{\text{GW}} \propto f^{2/p}$ where $p$ is given as in \eqref{eq:p}. Taking $\alpha=10^{14}$, $H = 10^{13}$ GeV and $\Omega_r^{\text{end}} = 10^{-10}$, we probe the effects of coupling on the spectrum of primordial GWs and on observability in Figure \ref{fig:gwobs_t}. We observe a boost in the GW spectrum during kination with a varying slope for all considered couplings. The spectrum after reheating and during hyperkination is flat. As the coupling order is increased, the constant energy density during reheating is enhanced. The hyperkinationary era prevents the enhancement in the spectrum from violating the BBN bound in this case as well. The spectrum falls more into an observable regime with decreasing orders of coupling. 

The freedom over choosing $t$ allows us to probe a wider class of theories as high as $\chi^2 R^2$ and their probable detection in the future. We probe the detectability for a few different parameter values for $H$ and $\Omega_r^{\text{end}}$ from the available parameter space in Figure \ref{fig:gwobs_allt}. We can see an array of effects from this variation. All cases result in giving a boosted spectrum with a plateaued hyperkination phase. The plateau during reheating also gains energy as we decrease the order of the coupling. The plateau during reheating and the boost during kination fall under the detectability of a number of future GW experiments.

\subsection{Minimal Starobinsky Model}

\begin{figure}[]
    \centering
    \includegraphics[width=\textwidth]{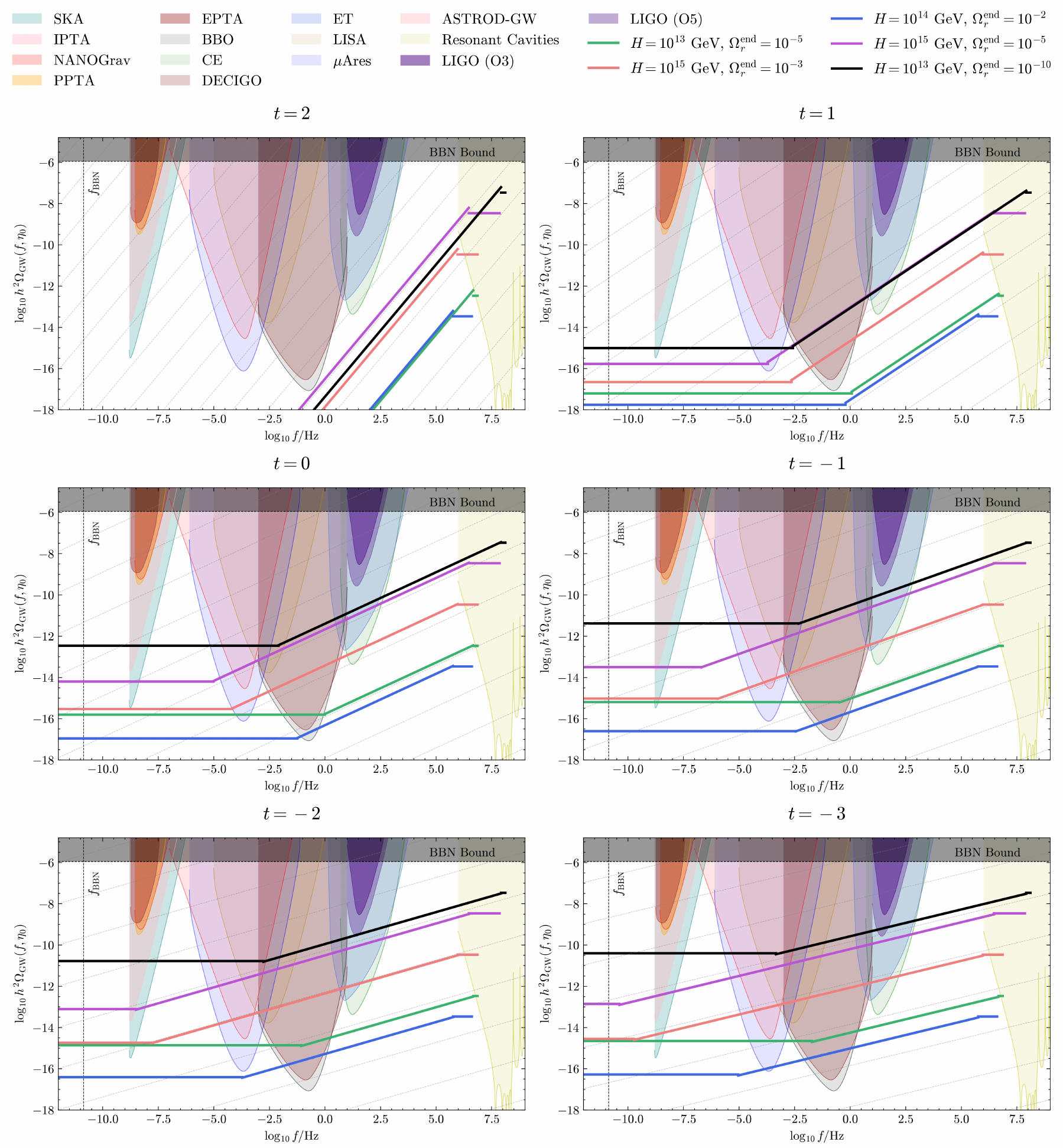}
    \caption{Spectrum of primordial GW for $\mathcal{L} \propto h(\chi)(R + \alpha R^2)$, where $h(\chi) = (1 + \xi \chi /M_{\text{Pl}})^t$, for a variety of coupling orders $t$. A boost in the GW spectrum during kination is of the form $\Omega_{\text{GW}} \propto f^{2/p}$. All cases result in giving a boosted spectrum with a crucial plateaued hyperkination phase. The plateau during reheating also gains energy as we decrease the order of the coupling. The gray region above highlights the BBN bound, while the vertical dashed line shows the frequency of BBN. The diagonal dashed gray lines show the boost, corresponding and different for each coupling, for different values of $H$ and $\Omega_r^{\text{end}}$. We considered $\alpha=10^{14}$.}
    \label{fig:gwobs_allt}
\end{figure}

One of the earliest models of inflation was the Starobinsky model \cite{Starobinsky:1980te}, $\mathcal{L} \propto R + \alpha R^2$, which we can retrieve from our model by setting $t=0$. The observability of this accepted theory is also tested in this work. The centre-left panel in Figure \ref{fig:gwobs_allt} shows the GW observability for the coupling $t=0$ or the minimal Starobinsky model in Palatini formalism. The bottom-left panel in Figure \ref{fig:paramspace} shows parameter space for which the theory remains valid. The availability of a large parameter space, the overlap of the GW spectra with observable region and a wide acceptance of this theory makes it exciting to anticipate future observations.

The observation of the plateau during reheating will constrain the values of $\Omega_r^{\text{end}}$ and $H$ if such is observed in the near future. However, to confirm a particular theory, the observation of boost during kination is required, which will give the spectral shape of the enhancement with respect to frequency. However, the plateau from hyperkination is outside any observable region of \textit{current-future} detectors. This prompts a need for GW detectors in the kHz to GHz frequency range, which can confirm the presence of the boost from kination and the bound from hyperkination. One proposal to detect ultra-high frequency gravitational waves (UHF-GWs), using \textit{inverse Gertsenshtein eﬀect} \cite{gertsenshtein1962wave}, has gained popularity recently. This has placed resonant cavities (RC) as promising candidates for detecting gravitational wave signals from these epochs within the MHz - GHz UHF-GW band \cite{Aggarwal_2021}.

\section{Summary and Conclusions} \label{sec:conclusion}

In this paper, we focused on the possibility of observing primordial gravitational waves produced during the eras of hyperkination, kination, and reheating. We considered a model with a higher order curvature term coupled to the inflaton field of the form $h(\chi) = (1+ \xi \chi /M_{\text{Pl}})^t$, where we can adjust the order of the coupling $t$. This model features a non-oscillatory inflaton potential that allows the inflaton to roll freely after it comes down from the inflationary plateau. The presence of $R^2$ terms in the Einstein frame results in non-canonical kinetic terms that are responsible for hyperkination and the flat and low potential demands movement of the scalar field under kinetic domination from canonical kinetic terms, giving the epoch of kination. We probe a wide class of theories from lower-order $(t<0)$, Starobinsky ($t=0$), and non-minimal coupling between the inflaton field and $R^2$ ($t>0$). We formulated the dynamics of the field during those times and obtained a bound on the length of the phase of hyperkination, $\Delta N_{\text{hyp}} \leq 0.2$, and the order of coupling, $t \leq 2$, from the requirement of the field derivative to be a kinetic attractor which restricts up to $\chi^2 R^2$. However, such a model for $t>0$ does not naturally give out reheating. We fixed this issue by providing a supplementary mechanism that governs the decay of the scalar field and brings out the conditions necessary for radiation domination and, thus, reheating with $T_{\text{reh}} \sim 10^8$ GeV. We assumed that the history of the universe is as follows: inflation, hyperkination, kination, reheating (BBN) followed by radiation and matter domination under $\Lambda$CDM.

We then solved the evolution of the scale factor with time during this history. The change in scale factor due to these new epochs brings out changes in tensor mode perturbations and, thus, in the observed GW spectrum. We solved the Mukhanov-Sasaki equation to obtain the perturbations and observed that as soon as the modes enter the kination phase, they obtain a large amount of energy. We also observed that coupling of lesser order boosts the perturbations during hyperkination and kination and, having a smaller duration of hyperkination, dilutes the contribution of GWs from kination. Following the mode calculation, we examined the spectral energy density coming from those primordial GWs. The spectrum features two plateaus, one after inflation during hyperkination and the other during reheating. The two plateaus are connected by the boost, $\Omega_{\text{GW}} \propto f^{2/p}$, gained during the phase of kination. The enhancement changes as a function of the coupling $t$. The epoch of hyperkination prevents the boost in the spectrum from violating the BBN bounds, thus demanding the presence of $R^2$ terms in the Einstein-Hilbert action whenever a runaway-like, non-oscillatory potential is considered. We probed the available parameter space for $\alpha$, $H$, and $\Omega_{r}^{\text{end}}$ for different orders of coupling, which are required to check the validity of the theory. Finally, we checked the observability of these modified spectra by current and future GW detectors by superimposing the $\Omega_{\text{GW}}$ spectra against the power-law integrated sensitivity curves. We find that as we decrease the order of the coupling, the spectra shift towards a more observable regime. We also highlighted the future observability of GWs from minimal Starobinsky model in Palatini formalism. The observation of the plateau during reheating will constrain the $H$ and $\Omega_{r}^{\text{end}}$ values, while the spectral shape of the boost obtained during kination will confirm the nature of the theory. To probe the bound placed by hyperkination that prevents the boost from kination from violating the Big Bang Nucleosynthesis, GW detector experiments, like detection via RC in the MHz-GHz frequency regime, are needed. The flat spectrum from reheating for lower-order coupling ($t<0$) can be matched with PTA and NANOGrav observations of the SGWB to constrain the parameter space. We can also place bounds on the parameter space detectable by future experiments; we leave these crucial analyses for the future.

Observations of primordial GWs will have large implications on early universe physics. A confirmation of this nature will prove the existence of higher-order curvature terms and the presence of non-canonical kinetic terms in Palatini gravity. It will supplement the theory of cosmic inflation by proving another of its predictions. Future GW observation holds a lot of potential for understanding the fundamental nature of physics and our universe.

\acknowledgments
HJK is funded by the INSPIRE Scholarship for Higher Education, Department of Science and Technology (DST), Government of India. The authors acknowledge the use of the following packages for this work: \texttt{NumPy} \cite{NumPy}, \texttt{Matplotlib} \cite{Matplotlib}, \texttt{Pandas} \cite{Pandas,Pandas2}, \texttt{SciPy} \cite{SciPy}, and \texttt{Mathematica} \cite{Mathematica}.  

\appendix
\section{Derivation for $\Bar{\Bar{\chi}}$} \label{app:chi}

We start with \eqref{eq:chi_relation}, taking its derivative with respect to time will give,
\begin{align}
    \Ddot{\chi} 
    & = \frac{1}{2} \left( \frac{1}{3 \alpha \kappa h} \frac{6 M_{\text{Pl}}^2 - \Bar{\chi}^2}{ \Bar{\chi}^2} \right)^{-1/2} \cdot \frac{1}{3 \alpha \kappa} \frac{\mathrm{d}}{\mathrm{d} \tau} \left[  \frac{1}{h} \left( \frac{6 M_{\text{Pl}}^2}{\Bar{\chi}^2} - 1 \right) \right]\\
    \Ddot{\chi} 
    & = \frac{H}{6 \alpha \kappa \Dot{\chi}} \left[ -\frac{h'}{h^2} \Bar{\chi} \left( \frac{6 M_{\text{Pl}}^2}{\Bar{\chi}^2} - 1 \right) - \frac{12 \Bar{\Bar{\chi}} M_{\text{Pl}}^2 }{h \Bar{\chi}^3} \right]
\end{align}
Note that the dot represents the derivative with respect to time $\tau$, and the overhead bars denote the derivative with respect to a number of $e$-folds $N$.

\begin{align}
    2[1+3\alpha \kappa h \Dot{\chi}^2] &= 2 + 3 \alpha \kappa h \frac{2}{3 \alpha \kappa h} \frac{6 M_{\text{Pl}}^2 - \Bar{\chi}^2}{ \Bar{\chi}^2} = \frac{12 M_{\text{Pl}}^2}{\Bar{\chi}^2}\\
    6H \left[ 1 + \alpha\kappa  h \Dot{\chi}^2 \right] \Dot{\chi} &= 6H \left[ 1 + \alpha \kappa h \frac{1}{3 \alpha \kappa h} \frac{6 M_{\text{Pl}}^2 - \Bar{\chi}^2}{ \Bar{\chi}^2} \right] \Dot{\chi} = 4 H \left[ \frac{\Bar{\chi}^2 + 3 M_{\text{Pl}}^2 }{ \Bar{\chi}^2} \right] \Dot{\chi}\\
    \frac{3 \alpha \kappa}{2} h' \Dot{\chi}^4 
    &= \frac{3 \alpha \kappa}{2} h' H \Bar{\chi} \Dot{\chi}^3
\end{align}
Combining all the terms,
we get,
\begin{align}
    &2[1+3\alpha \kappa h \Dot{\chi}^2] \Ddot{\chi} + 6H \left[ 1 + \alpha \kappa h \Dot{\chi}^2 \right] \Dot{\chi} + \frac{3 \alpha \kappa}{2} h' \Dot{\chi}^4 = \frac{12 M_{\text{Pl}}^2}{\Bar{\chi}^2} \Ddot{\chi} + 4 H \left[ \frac{\Bar{\chi}^2 + 3 M_{\text{Pl}}^2 }{ \Bar{\chi}^2} \right] \Dot{\chi} + \frac{3 \alpha \kappa}{2} h' H \Bar{\chi} \Dot{\chi}^3\\
    0&= 2 \left[ -\frac{h'}{h} \Bar{\chi} \left( 6 M_{\text{Pl}}^2 - \Bar{\chi}^2 \right) M_{\text{Pl}}^2 - \frac{12 \Bar{\Bar{\chi}}}{ \Bar{\chi}} M_{\text{Pl}}^4 \right] + \frac{4}{3} \left( \Bar{\chi}^2 + 3 M_{\text{Pl}}^2 \right)(6 M_{\text{Pl}}^2 - \Bar{\chi}^2) +   \frac{h' \Bar{\chi}}{6 h}(6 M_{\text{Pl}}^2 - \Bar{\chi}^2)^2\\
    0&= (6 M_{\text{Pl}}^2 - \Bar{\chi}^2) \left[ -\frac{h' \Bar{\chi}}{h} M_{\text{Pl}}^2 + \frac{4}{3} \left( \Bar{\chi}^2 + 3 M_{\text{Pl}}^2 \right) - \frac{h' \Bar{\chi}^3}{6 h }\right] - \frac{24 M_{\text{Pl}}^4 \Bar{\Bar{\chi}}}{ \Bar{\chi}}
\end{align}
\begin{equation}
    \Bar{\Bar{\chi}} = \frac{\Bar{\chi}}{24 M_{\text{Pl}}^4} (6 M_{\text{Pl}}^2 - \Bar{\chi}^2) \left[ -\frac{h'}{h} \left( 1 + \frac{\Bar{\chi}^2}{6 M_{\text{Pl}}^2} \right) \Bar{\chi} M_{\text{Pl}}^2 + \frac{4}{3} \left( \Bar{\chi}^2 + 3 M_{\text{Pl}}^2 \right) \right] 
\end{equation}

\section{$\chi^2 R^2$ Inflation} \label{app:inflation}

For, $V = \beta \chi^2$ and $h = \gamma \chi^2$ we obtain the following parameters for this case,
\begin{equation}
    \epsilon_\text{v} = \frac{2}{\chi^2} \hspace{0.3cm} \text{and} \hspace{0.3cm} |\eta_\text{v}| =  \frac{6}{\chi^2}
\end{equation}
\begin{equation}
    n_s = 1 - \frac{24}{\chi^2} \hspace{0.3cm} \text{and} \hspace{0.3cm} r = \frac{32}{\chi^2}
\end{equation}
\begin{equation}
    U(\chi) = \frac{\beta }{\gamma  \chi ^2 (8 \alpha \kappa \beta +\gamma )} \hspace{0.3cm} \text{and} \hspace{0.3cm} \chi(\chi) = \frac{\ln{\chi}}{\sqrt{\gamma + 8 \alpha \kappa \beta}}
\end{equation}

\begin{figure}
    \centering
    \includegraphics[width=\textwidth]{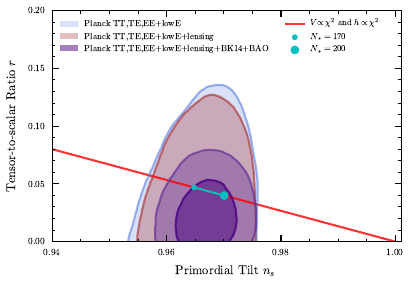}
    \caption{Tensor-to-scalar ratio $r$ plotted against the primordial tilt $n_s$ with Planck observations \cite{Collaboration_Akrami_Arroja_Ashdown_Aumont_Baccigalupi_Ballardini_Banday_Barreiro_Bartolo_2019b} for $V \propto \chi^2$ and $h \propto \chi^2$.}
    \label{fig:phi2R2}
\end{figure}
Figure \ref{fig:phi2R2} shows the Tensor-to-scalar ratio $r$ plotted against the primordial tilt $n_s$ with Planck observations for the considered case. Thus we find this model as valid and claim that successful inflation is observed in $\chi^2 R^2$ theory with a quadratic potential.

\section{Derivation for Scale Factors} \label{app:scalefactor}

At $\eta = \eta_{\text{end}}$, from the normalization condition,
\begin{equation}
    a(\eta_{\text{end}}) = \left[ \frac{-1}{(1-\epsilon)H \eta_{\text{end}}} \right]^{1/(1-\epsilon)} = 1 \implies H \eta_{\text{end}} = -1 \because \epsilon = 0
\end{equation}

For the epoch of hyperkination, which is right after the end of inflation, we consider its energy density as given in \eqref{eq:rho-chi}, where $\Bar{\chi}$ goes as the first branch in \eqref{eq:transition}, and the initial condition is written in terms of $\Delta N_{\text{hyp}}$ from \eqref{eq:deltaN}. We get,
\begin{equation}
    \rho = \frac{1 - e^{2(N - \Delta N_{\text{hyp}})} }{6 \alpha \kappa h ~e^{4(N - \Delta N_{\text{hyp}})}} = \frac{1 - a^2 e^{- 2\Delta N_{\text{hyp}}} }{6 \alpha \kappa h ~a^4 e^{-4\Delta N_{\text{hyp}}}}
\end{equation}
To get rid of $\alpha$, we evaluate $\rho$ at the end of inflation, at $N = N_{\text{inf}}^{\text{end}} = 0$,
\begin{equation}
    \rho_{\text{end}} = \frac{1 - e^{- 2\Delta N_{\text{hyp}}} }{6 \alpha \kappa h ~e^{-4\Delta N_{\text{hyp}}}} = 3 H^2 \implies \frac{1 - e^{- 2\Delta N_{\text{hyp}}} }{18 \kappa h H^2 ~e^{-4\Delta N_{\text{hyp}}}} = \alpha
\end{equation}

\begin{align}
    \mathrm{d} \eta & =  \mathrm{d} a \sqrt{\frac{18 \alpha \kappa h ~e^{-4 \Delta N_{\text{hyp}}}}{1 - a^2 e^{- 2\Delta N_{\text{hyp}}} }}\\
    \implies \eta & = \frac{\sqrt{e^{2 \Delta N_{\text{hyp}} } - 1} \arcsin\left(e^{- \Delta N_{\text{hyp}}} a\right)}{H} + C
\end{align}
where $C$ is the constant of integration. Demanding continuity of $a$ at $\eta = \eta_{\text{end}}$,
\begin{equation}
   C = \eta_{\text{end}} - \frac{\sqrt{e^{2 \Delta N_{\text{hyp}} } - 1} \arcsin\left(e^{- \Delta N_{\text{hyp}}} \right)}{H}
\end{equation}
Solving for $a$ gives us, 
\begin{align}
    a (\eta) & = e^{\Delta N_{\text{hyp}}} \sin{ \left( \frac{H \eta + 1}{\sqrt{e^{2 \Delta N_{\text{hyp}} } - 1}} + \arcsin\left(e^{- \Delta N_{\text{hyp}}} \right) \right)}
\end{align}
where the last equality uses the fact that $H\eta_{\text{end}}=-1$. 

Now, we move towards the epoch of standard kination. For our non-minimal coupling, we get a non-standard scaling of $\rho$ during this phase, along with the corrections for the decay factor $\Gamma$ from reheating. We have
\begin{equation}
    \rho = \frac{c}{3 \alpha \kappa h} a^{t \sqrt{6} - 6} a^{-\Gamma \tau} = \frac{6c H^2 ~e^{-4\Delta N_{\text{hyp}}}}{1 - e^{- 2\Delta N_{\text{hyp}}} } \cdot a^{t \sqrt{6} - 6 -\Gamma \tau} \equiv B a^{t \sqrt{6} - 6-\Gamma \tau}
\end{equation}
\begin{align}
    \mathrm{d} \eta & =  \frac{\mathrm{d} a}{a^2} \sqrt{\frac{3}{B a^{t \sqrt{6} - 6-\Gamma\tau} }} 
    \implies \eta = \frac{2 \sqrt{3}}{\sqrt{B} \left(\Gamma\tau -t\sqrt{6} + 4\right) a^{\frac{t\sqrt{6} - 4-\Gamma\tau}{2}}} + C 
\end{align}
Let, $\frac{\Gamma\tau +4 - t\sqrt{6}}{2} \equiv p$, we now demand the continuity at $\eta = \eta_{\text{kin}}$,
\begin{equation}
    C = \eta_{\text{kin}} - \frac{a_{\text{kin}}^{p} \sqrt{3}}{\sqrt{B} p} 
\end{equation}
where 
\begin{equation}
    a_{\text{kin}} \equiv a(\eta_{\text{kin}}) = e^{\Delta N_{\text{hyp}}} \sin{ \left( \frac{H \eta_{\text{kin}} + 1}{\sqrt{e^{2 \Delta N_{\text{hyp}} } - 1}} + \arcsin\left(e^{- \Delta N_{\text{hyp}}} \right) \right)}
\end{equation}
therefore, we have,
\begin{equation}
    a(\eta) = \left( p \frac{H e^{-\Delta N_{\text{hyp}}}\sqrt{2c}}{\sqrt{e^{2\Delta N_{\text{hyp}}} -1 }} (\eta - \eta_{\text{kin}}) +  a_{\text{kin}}^{p} \right)^{1/p}
\end{equation}

Kination is followed by the era of radiation domination or reheating. For this epoch, the energy scale is $a^{-4}$. We have
\begin{equation}
    \rho = 3H^2a^{-4}
\end{equation}
\begin{equation}
    \mathrm{d} \eta = \frac{\mathrm{d}a}{a^2} \sqrt{\frac{3}{3 H^2 a^{-4}}} = \frac{\mathrm{d}a}{H} \implies \eta = \frac{a}{H} + C
\end{equation}
Imposing the condition of continuity at $\eta = \eta_{\text{reh}}$,
\begin{equation}
    C = \eta_{\text{reh}} - \frac{a_{\text{reh}}}{H}
\end{equation}
where,
\begin{equation}
    a_{\text{reh}} \equiv a(\eta_{\text{reh}}) = \left( p \frac{H e^{-\Delta N_{\text{hyp}}}\sqrt{2c}}{\sqrt{e^{2\Delta N_{\text{hyp}}} -1 }} (\eta_\text{reh} - \eta_{\text{kin}}) +  a_{\text{kin}}^{p} \right)^{1/p}
\end{equation}
Therefore we get,
\begin{equation}
    a(\eta) = H(\eta - \eta_{\text{reh}}) + a_{\text{reh}}
\end{equation}

\section{Solving for Mode Functions} \label{app:modefn}

We have the most general solution for the modes as given in \eqref{eq:MS_mostgensol},
\begin{equation} \label{eq:MS_mostgensol2}
    v(x) = \sqrt{x} \left( c_1 H_\nu^{(1)}(x) + c_2 H_\nu^{(2)}(x) \right)
\end{equation}
We start with the inflationary era and work in the sub-Hubble limit. The Bunch-Davies vacuum condition requires that the solution $v(x)$ behaves like a positive frequency mode in the distant past (as $x\to \infty$ or $\eta \to -\infty$). In this limit, the Hankel functions $H_\nu^{(1)}(x)$ and $H_\nu^{(2)}(x)$ behave as incoming and outgoing waves respectively,
\begin{align} \label{eq:hankel_largelimit}
    H_\nu^{(1)}(x) \sim \sqrt{\frac{2}{\pi x}} e^{i\left( x - \frac{\nu \pi}{2} - \frac{\pi}{4} \right)} \hspace{0.3cm} \text{and} \hspace{0.3cm} H_\nu^{(2)}(x) \sim \sqrt{\frac{2}{\pi x}} e^{-i\left( x - \frac{\nu \pi}{2} - \frac{\pi}{4} \right)} 
\end{align}
$H_\nu^{(1)}(x)$ represents an outgoing wave (positive frequency mode), as it oscillates with a positive exponential phase. $H_\nu^{(2)}(x)$ represents an incoming wave (negative frequency mode), as it oscillates with a negative exponential phase. Both decay as $\sim 1/\sqrt{x}$ as $x \to \infty$, which ensures a decaying envelope for large arguments. During this phase, the mode functions should obey the Bunch-Davies vacuum condition as 
\begin{equation}
    \lim_{\eta \to -\infty} v(\eta,k) = \frac{1}{\sqrt{2k}} e^{-ik\eta}
\end{equation}
For the Bunch-Davies vacuum, we require that only the positive frequency mode survives in the limit $x \to \infty$, which means we must set $c_2 = 0$. This gives us,
\begin{equation}
    v(x) = c_1 \sqrt{x} H_{\nu_I}^{(1)}(x)
\end{equation}
In the sub-Hubble regime ($x\to\infty$), the Hankel function of the first kind takes the form of the first equation in \eqref{eq:hankel_largelimit}. Thus for $x \gg 1$, the mode function becomes,
\begin{equation}
    v(x) \sim c_1 \sqrt{\frac{2}{\pi}} e^{i\left( x - \frac{\nu_I \pi}{2} - \frac{\pi}{4} \right)}
\end{equation}
The Bunch-Davies vacuum condition in the sub-Hubble regime should match the Minkowski vacuum, which corresponds to a simple plane wave in the distant past,
\begin{equation}
    v(x) \sim \frac{1}{\sqrt{2 k}} e^{ix}
\end{equation}
This is the positive-frequency mode of the Minkowski vacuum in the sub-Hubble regime. To ensure that $v(x)$ matches this, we need to set $c_1$ such that,
\begin{equation}
    c_1 \sqrt{\frac{2}{\pi}} = \frac{1}{\sqrt{2 k}} e^{\frac{i\pi}{4}(1+2 \nu_I)} \implies c_1 = \sqrt{\frac{\pi}{4k}} e^{\frac{i\pi}{4}(1+2 \nu_I)}
\end{equation}
Thus, the solution for $v(x)$ that satisfies the Bunch-Davies vacuum condition during inflation in the sub-Hubble regime is
\begin{align}
    \begin{split}
        v(x) &= \sqrt{\frac{\pi}{4k}} \sqrt{x}~ e^{\frac{i\pi}{4}(1+2 \nu_I)} H_{\nu_I}^{(1)}(x)
    \end{split}\\
    \begin{split} \label{eq:modefn_sol1}
        v_k^s(\eta) & = \sqrt{\frac{\pi}{4k}} \sqrt{-k \eta} ~e^{\frac{i\pi}{4}(1+2 \nu_I)} H_{\nu_I}^{(1)}(-k \eta)
    \end{split}
\end{align}
where $\nu_I = 3/2 + \epsilon$. Over their cosmic evolution, the modes
stretch and exit the Hubble radius, evolving beyond Bunch-Davies vacuum solutions. After inflation, they re-enter the Hubble radius, this time following the general sub-Hubble form. For the eras of hyperkination, $\nu=-1/2$, in the sub-Hubble regime ($x = k \eta \gg 1$), we get the full solution, given as,
\begin{equation}
    v_k^s(\eta) = \frac{1}{\sqrt{2k}} \left[ \alpha_+(k) e^{-ik\eta} + \alpha_-(k) e^{ik\eta}\right]
\end{equation}
A similar solution with different coefficients ($\gamma_\pm$) follows for the era of radiation-domination. We now focus on the era of kination. This phase does not require a Bunch-Davies vacuum condition and thus can have negative-frequency modes. Therefore the solution is similar to \eqref{eq:modefn_sol1},
\begin{align}
    \begin{split} \label{eq:modefn_sol2}
        v_k^s(\eta) &= \sqrt{\frac{\pi}{4k}} \sqrt{kz} ~e^{\frac{i\pi}{4}(1+2 \nu_{II})} H_{\nu_{II}}^{(1)}(kz) + \sqrt{\frac{\pi}{4k}} \sqrt{kz} ~e^{-\frac{i\pi}{4}(1+2 \nu_{II})} H_{\nu_{II}}^{(2)}(kz) 
    \end{split}\\
    \begin{split}
        & = \sqrt{\frac{\pi z}{4}} \left[ \beta_+(k) ~e^{-\frac{i\pi}{4}(1+2 \nu_{II})} H_{\nu_{II}}^{(2)}(kz) + \beta_-(k) ~e^{\frac{i\pi}{4}(1+2 \nu_{II})} H_{\nu_{II}}^{(1)}(kz)\right]
    \end{split}
\end{align}
where $\nu_{II} = 1/2 - 1/p$ and the overall constant and phase have been chosen such that the mode functions have a simple sub-Hubble
($x \gg 1$) limit.

\section{Matching the Mode Functions} \label{app:modematching}

We start with the transition from inflation to hyperkination, which corresponds to $\eta = \eta_{\text{end}}$. During hyperkination, the Mukhanov-Sasaki equation and its solution reads as,
\begin{equation}
    v^{s''}_k + k^2 v^s_k = 0 \implies v^s_k (\eta) = \frac{1}{\sqrt{2k}} \left[ \alpha_+(k) e^{-ik\eta} + \alpha_-(k) e^{ik\eta}\right]
\end{equation}
We match this with the standard slow roll result for inflation at $\eta_{\text{end}}$,
\begin{equation} \label{eq:modefn_match1}
    \sqrt{\frac{\pi}{2}} \sqrt{x_{\text{end}}} ~e^{\frac{i\pi}{4}(1+2 \nu_I)} H_{\nu_I}^{(1)}(x_{\text{end}}) =   \alpha_+ e^{ik |\eta_{\text{end}}|} + \alpha_- e^{-ik |\eta_{\text{end}}|}
\end{equation}
where $x_{\text{end}} \equiv k|\eta_{\text{end}}|$. Matching the derivatives gives,
\begin{equation} \label{eq:modefn_match2}
    i e^{\frac{i\pi}{4}(1+2 \nu_I)} \sqrt{\frac{\pi}{2}} \left[ \frac{1}{\sqrt{x_{\text{end}}} } \left( \frac{1}{2} + \nu \right) H_{\nu_I}^{(1)}(x_{\text{end}}) - \sqrt{x_{\text{end}}} H_{\nu_I+1}^{(1)}(x_{\text{end}}) \right] = - \alpha_+ e^{ik |\eta_{\text{end}}|} + \alpha_- e^{-ik |\eta_{\text{end}}|}
\end{equation}
Adding  and subtracting \eqref{eq:modefn_match1} and \eqref{eq:modefn_match2} gives us,
\begin{equation}
    \alpha_\mp = \frac{1}{2} e^{\frac{i\pi}{4}(1+2 \nu_I) \pm i x_{\text{end}}} \sqrt{\frac{\pi}{2}} \left[ H_{\nu_I}^{(1)}(x_{\text{end}}) \left( \sqrt{x_{\text{end}}} \pm i \frac{1}{\sqrt{x_{\text{end}}} } \left( \frac{1}{2} + \nu \right) \right) \mp i \sqrt{x_{\text{end}}} H_{\nu_I+1}^{(1)}(x_{\text{end}}) \right]
\end{equation}
In the super-Hubble limit, $x_{\text{end}} \ll 1$,
\begin{align}
    \alpha_\mp &= \frac{1}{2} e^{\frac{i\pi}{4}(1+2 \nu_I)} \sqrt{\frac{\pi}{2}} \left[ H_{\nu_I}^{(1)}(x_{\text{end}}) \left( \pm  \frac{i}{\sqrt{x_{\text{end}}} } \left( \frac{1}{2} + \nu \right) \right) \mp i \sqrt{x_{\text{end}}} H_{\nu_I+1}^{(1)}(x_{\text{end}}) \right]
\end{align}
This gives,
\begin{equation}
    \alpha_\mp = \pm 2^{\nu_I - 1} e^{\frac{i\pi}{4}(1+2 \nu_I)} \sqrt{\frac{1}{2 \pi}}  \left( \frac{1}{2} - \nu \right) \Gamma(\nu_I)  \frac{1}{x_{\text{end}}^{\nu_I + 1/2}}
\end{equation}
Using $\nu_I = 3/2 + \epsilon$, this expression gives,
\begin{align}
    \alpha_\mp &= \pm 2^{\epsilon-1} e^{\frac{i\pi}{2}\epsilon}   (\epsilon + 1) \frac{\Gamma ( 3/2 + \epsilon )}{\Gamma ( 3/2 )} \left( \frac{H}{k} \right)^{2 + \epsilon}
\end{align}
where $\Gamma ( 3/2 ) = \sqrt{\pi}/2$ and $\eta_{\text{end}} = 1/H$. In pure de Sitter limit, with $\epsilon \to 0$, we get,
\begin{equation}
    \alpha_\mp = \pm \frac{H^2}{2k^2} 
\end{equation}

The next transition occurs at $\eta_{\text{kin}}$, for the phase moving from hyperkination to kination. Solving the Mukhanov-Sasaki equation as done in Appendix \ref{app:modefn}, we use \eqref{eq:modefn_sol2} with the redefinition of $y\equiv kz$ and $\nu_{II} = 1/2 - 1/p$.
\begin{equation} \label{eq:modefn_match6}
    v_k^s(\eta) = \sqrt{\frac{\pi}{4k}} \sqrt{y} \left[ \beta_- (k)~e^{\frac{i\pi}{4}(1+2 \nu_{II})} H_{\nu_{II}}^{(1)}(y) + \beta_+ (k) ~e^{-\frac{i\pi}{4}(1+2 \nu_{II})} H_{\nu_{II}}^{(2)}(y) \right]
\end{equation}
In the redefined variables $\eta_{\text{kin}}$, considering $\eta_{\text{kin}} \gg \eta_{\text{end}} = 1/H$ becomes
\begin{equation}
    y_{\text{kin}} \equiv k z_{\text{kin}} = k \left( \frac{\eta_{\text{kin}}}{2} + \frac{1}{H} \right) \approx \frac{k \eta_{\text{kin}}}{2}
\end{equation}
We redefine,
\begin{equation}
    r \equiv \sqrt{\frac{\pi}{2}}~e^{\frac{i\pi}{4}(1+2 \nu_{II})} \quad \text{and} \quad r^\ast \equiv \sqrt{\frac{\pi}{2}}~e^{\frac{-i\pi}{4}(1+2 \nu_{II})}
\end{equation}
Matching the mode functions at $y_{\text{kin}}$/$\eta_{\text{kin}}$,
\begin{equation} \label{eq:modefn_match3}
      \alpha_+ e^{-ik\eta_{\text{kin}}} + \alpha_- e^{ik\eta_{\text{kin}}} = \sqrt{y_{\text{kin}}} \left[ \beta_- r H_{\nu_{II}}^{(1)}(y_{\text{kin}}) + \beta_+ r^\ast H_{\nu_{II}}^{(2)}(y_{\text{kin}}) \right]
\end{equation}
Matching their derivatives,
\begin{equation}
    \begin{split}
        i\left( -\alpha_+ e^{-ik\eta_{\text{kin}}} + \alpha_- e^{ik\eta_{\text{kin}}} \right) 
        &= \frac{1}{2\sqrt{y_{\text{kin}}}} \left[ \beta_- r H_{\nu_{II}}^{(1)}(y_{\text{kin}}) + \beta_+ r^\ast H_{\nu_{II}}^{(2)}(y_{\text{kin}}) \right] \\
        & + \sqrt{y_{\text{kin}}} \left[ \beta_- r \frac{\mathrm{d} H_{\nu_{II}}^{(1)}}{\mathrm{d}y}  (y_{\text{kin}}) + \beta_+ r^\ast \frac{\mathrm{d} H_{\nu_{II}}^{(2)}}{\mathrm{d}y} (y_{\text{kin}}) \right]
    \end{split}
\end{equation}
Using \eqref{eq:modefn_match3},
\begin{equation} \label{eq:modefn_match4}
    \begin{split}
       \alpha_+ \left( -i -\frac{1}{2y_{\text{kin}}} \right) e^{-ik\eta_{\text{kin}}} + \alpha_- \left( i -\frac{1}{2y_{\text{kin}}} \right) e^{ik\eta_{\text{kin}}} 
        &= \sqrt{y_{\text{kin}}} \left[ \beta_- r \frac{\mathrm{d} H_{\nu_{II}}^{(1)}}{\mathrm{d}y}  (y_{\text{kin}}) + \beta_+ r^\ast \frac{\mathrm{d} H_{\nu_{II}}^{(2)}}{\mathrm{d}y} (y_{\text{kin}}) \right]
    \end{split}
\end{equation}
To get $\beta_-$, we multiply \eqref{eq:modefn_match4} by $H_{\nu_{II}}^{(2)}(y_{\text{kin}})$ and \eqref{eq:modefn_match3} by $\mathrm{d} H_{\nu_{II}}^{(2)} / \mathrm{d}y$ and subtract the latter from the former and use the Wronskian of the Hankel functions.
\begin{equation}
    \begin{split}
       H_{\nu_{II}}^{(2)} \left( \alpha_+ \left( -i -\frac{1}{2y_{\text{kin}}} \right) e^{-ik\eta_{\text{kin}}} + \alpha_- \left( i -\frac{1}{2y_{\text{kin}}} \right) e^{ik\eta_{\text{kin}}} \right)
        &= \sqrt{y_{\text{kin}}} \left[ \beta_- r H_{\nu_{II}}^{(2)} \frac{\mathrm{d} H_{\nu_{II}}^{(1)}}{\mathrm{d}y}  (y_{\text{kin}}) \right.\\
        & \left. + \beta_+ r^\ast H_{\nu_{II}}^{(2)} \frac{\mathrm{d} H_{\nu_{II}}^{(2)}}{\mathrm{d}y} (y_{\text{kin}}) \right]
    \end{split}
\end{equation}
\begin{equation}
    \begin{split}
        \beta_- 
        &= e^{-\frac{i\pi}{4}(1+2 \nu_{II})} \frac{\sqrt{\pi y_{\text{kin}}}}{i2\sqrt{2}} \left[ H_{\nu_{II}}^{(2)} 
        \left( \alpha_+ \left[ -i -\frac{1}{2y_{\text{kin}}} \right] e^{-ik\eta_{\text{kin}}} + \alpha_- \left[ i -\frac{1}{2y_{\text{kin}}} \right] e^{ik\eta_{\text{kin}}} \right) \right.\\
        & \left. - \frac{1}{2} \left( H_{\nu_{II}-1}^{(2)}  - H_{\nu_{II}+1}^{(2)}  \right) \left( \alpha_+ e^{-ik\eta_{\text{kin}}} + \alpha_- e^{ik\eta_{\text{kin}}} \right) \right]
    \end{split}
\end{equation}
Now to get $\beta_+$, we multiply \eqref{eq:modefn_match4} by $H_{\nu_{II}}^{(1)}(y_{\text{kin}})$ and \eqref{eq:modefn_match3} by $\mathrm{d} H_{\nu_{II}}^{(1)} / \mathrm{d}y$ and follow the same procedure, we get,
\begin{equation}
    \begin{split}
        \beta_+ 
        &= -e^{\frac{i\pi}{4}(1+2 \nu_{II})} \frac{\sqrt{\pi y_{\text{kin}}}}{i2\sqrt{2}} \left[ H_{\nu_{II}}^{(1)} 
        \left( \alpha_+ \left[ -i -\frac{1}{2y_{\text{kin}}} \right] e^{-ik\eta_{\text{kin}}} + \alpha_- \left[ i -\frac{1}{2y_{\text{kin}}} \right] e^{ik\eta_{\text{kin}}} \right) \right.\\
        & \left. - \frac{1}{2} \left( H_{\nu_{II}-1}^{(1)}  - H_{\nu_{II}+1}^{(1)}  \right) \left( \alpha_+ e^{-ik\eta_{\text{kin}}} + \alpha_- e^{ik\eta_{\text{kin}}} \right) \right]
    \end{split}
\end{equation}
Note that, we have $\alpha_+ = - \alpha_-$, we can write,
\begin{equation}
    \begin{split}
        \beta_- 
        &= e^{-\frac{i\pi}{4}(1+2 \nu_{II})} \frac{\sqrt{\pi y_{\text{kin}}}}{i2\sqrt{2}} \left[ H_{\nu_{II}}^{(2)} 
        \left( i\alpha_- \left( e^{-ik\eta_{\text{kin}}} +  e^{ik\eta_{\text{kin}}} \right) + \frac{\alpha_-}{2 y_{\text{kin}}} \left( e^{-ik\eta_{\text{kin}}} -  e^{ik\eta_{\text{kin}}} \right) \right) \right.\\
        & \left. - \frac{\alpha_-}{2} \left( H_{\nu_{II}-1}^{(2)}  - H_{\nu_{II}+1}^{(2)}  \right) \left( - e^{-ik\eta_{\text{kin}}} +  e^{ik\eta_{\text{kin}}} \right) \right]
    \end{split}
\end{equation}
\begin{equation}
    \begin{split}
        \beta_- 
        &= e^{-\frac{i\pi}{4}(1+2 \nu_{II})} \sqrt{\frac{\pi y_{\text{kin}}}{2}} ~\alpha_- \left[ H_{\nu_{II}}^{(2)} 
        \left( \cos{(k\eta_{\text{kin}})} - \frac{1}{2y_{\text{kin}}} \sin{(k\eta_{\text{kin}})} \right) \right.\\
        & \left. -  \left( H_{\nu_{II}-1}^{(2)}  - H_{\nu_{II}+1}^{(2)}  \right) \sin{(k\eta_{\text{kin}})} \right]
    \end{split}
\end{equation}
Similarly,
\begin{equation}
    \begin{split}
        \beta_+ 
        &= -e^{\frac{i\pi}{4}(1+2 \nu_{II})} \sqrt{\frac{\pi y_{\text{kin}}}{2}} ~\alpha_- \left[ H_{\nu_{II}}^{(1)} 
        \left( \cos{(k\eta_{\text{kin}})} - \frac{1}{2y_{\text{kin}}} \sin{(k\eta_{\text{kin}})} \right) \right.\\
        & \left. -  \left( H_{\nu_{II}-1}^{(1)}  - H_{\nu_{II}+1}^{(1)}  \right) \sin{(k\eta_{\text{kin}})} \right]
    \end{split}
\end{equation}
The Hankel functions are evaluated at $y_{\text{kin}}$. We now take the super-Hubble limit $k \eta_{\text{kin}} \ll 1$ and taking $2 y_{\text{kin}} = k \eta_{\text{kin}}$,
\begin{align}
        \beta_- 
        &= e^{-\frac{i\pi}{4}(1+2 \nu_{II})} \sqrt{\frac{\pi y_{\text{kin}}}{2}} ~\alpha_- \left[ -  \left( H_{\nu_{II}-1}^{(2)}  - H_{\nu_{II}+1}^{(2)}  \right) k\eta_{\text{kin}} \right]
\end{align}
\begin{align}
        \beta_- 
        &= - e^{-\frac{i\pi}{4}(1+2 \nu_{II})} \sqrt{\frac{\pi}{4}} ~\frac{H^2}{2k^2} \cdot 2 \left. \frac{\mathrm{d} H_{\nu_{II}}^{(2)}}{\mathrm{d}y} \right|_{y_{\text{kin}}} (k\eta_{\text{kin}})^{3/2}\\
        \begin{split} \label{eq:modefn_match5}
            & = e^{-\frac{i\pi}{4}(1+2 \nu_{II})} \sqrt{\frac{\pi}{4}} ~\frac{H^2}{2k^2} \cdot 2 \frac{i}{\pi} \Gamma(\nu_{II}+1) \left( \frac{y_{\text{kin}}}{2} \right)^{-\nu_{II}-1} (k\eta_{\text{kin}})^{3/2}
        \end{split}
\end{align}
\begin{equation}
    \beta_- = ie^{-\frac{i\pi}{4}(1+2 \nu_{II})} 2^{2\nu_{II}}~\frac{H^2}{k^2}  \frac{\Gamma(3/2-1/p)}{\Gamma(3/2)}  (k\eta_{\text{kin}})^{1/p}
\end{equation}
Similarly,
\begin{align}
    \beta_+ 
        &= -e^{\frac{i\pi}{4}(1+2 \nu_{II})} \sqrt{\frac{\pi y_{\text{kin}}}{2}} ~\alpha_- \left[ -  \left( H_{\nu_{II}-1}^{(1)}  - H_{\nu_{II}+1}^{(1)}  \right) k\eta_{\text{kin}} \right]
\end{align}
\begin{equation}
    \beta_+ = ie^{\frac{i\pi}{4}(1+2 \nu_{II})} 2^{2\nu_{II}}~\frac{H^2}{k^2}  \frac{\Gamma(3/2-1/p)}{\Gamma(3/2)}  (k\eta_{\text{kin}})^{1/p}
\end{equation}
with the relation,
\begin{equation}
    \beta_+ = i \beta_-
\end{equation}

The last transition of our interest occurs when the fields go from kination to reheating at $\eta = \eta_{\text{reh}}$. The Mukhanov-Sasaki equation is identical to that of hyperkination, and thus we get the solutions,
\begin{equation}
    v^s_k (\eta) = \frac{1}{\sqrt{2k}} \left[ \gamma_+(k) e^{-ik\eta} + \gamma_-(k) e^{ik\eta}\right]
\end{equation}
Note that, in redefined variables, considering $\eta_{\text{reh}} \gg \eta_{\text{kin}} \gg |\eta_{\text{end}}|$,
\begin{equation}
    y_{\text{reh}} \equiv k z_{\text{reh}} = k \left( \eta_{\text{reh}} - \frac{\eta_{\text{kin}}}{2} + \frac{1}{H} \right) \approx k \eta_{\text{reh}}
\end{equation}
Matching \eqref{eq:modefn_match5} with \eqref{eq:modefn_match6} at $\eta_{\text{reh}}/y_{\text{reh}}$,
\begin{equation}
    \gamma_+ e^{-ik\eta_{\text{reh}}} + \gamma_- e^{ik\eta_{\text{reh}}} = \sqrt{y_{\text{reh}}} \left[ \beta_- r H_{\nu_{II}}^{(1)}(y_{\text{reh}}) + \beta_+ r^\ast H_{\nu_{II}}^{(2)}(y_{\text{reh}}) \right]
\end{equation}
and the derivatives,
\begin{equation}
    \begin{split}
        i\left( -\gamma_+ e^{-ik\eta_{\text{reh}}} + \gamma_- e^{ik\eta_{\text{reh}}} \right) 
        &= \frac{1}{2\sqrt{y_{\text{reh}}}} \left[ \beta_- r H_{\nu_{II}}^{(1)}(y_{\text{reh}}) + \beta_+ r^\ast H_{\nu_{II}}^{(2)}(y_{\text{reh}}) \right] \\
        & + \sqrt{y_{\text{reh}}} \left[ \beta_- r \frac{\mathrm{d} H_{\nu_{II}}^{(1)}}{\mathrm{d}y}  (y_{\text{reh}}) + \beta_+ r^\ast \frac{\mathrm{d} H_{\nu_{II}}^{(2)}}{\mathrm{d}y} (y_{\text{reh}}) \right]
    \end{split}
\end{equation}
Summing both the expressions and rearranging gives,
\begin{equation}
    \begin{split}
        \gamma_- 
        &= \frac{e^{-ik\eta_{\text{reh}}}}{2} \left\{ \beta_- r \left[  H_{\nu_{II}}^{(1)}(y_{\text{reh}}) \left( \sqrt{y_{\text{reh}}} - \frac{i}{2\sqrt{y_{\text{reh}}}} \right) -i\sqrt{y_{\text{reh}}} H_{\nu_{II}+1}^{(1)}(y_{\text{reh}}) \right] \right.\\
        & \left. + \beta_+ r^\ast \left[  H_{\nu_{II}}^{(2)}(y_{\text{reh}}) \left( \sqrt{y_{\text{reh}}} - \frac{i}{2\sqrt{y_{\text{reh}}}} \right) -i\sqrt{y_{\text{reh}}} H_{\nu_{II}+1}^{(2)}(y_{\text{reh}}) \right] \right\}
    \end{split}
\end{equation}
In the super-Hubble limit, $y_{\text{reh}} \ll 1$,
\begin{align}
    \gamma_-   &= - \frac{i}{4\sqrt{y_{\text{reh}}}}   \left( \beta_- r H_{\nu_{II}}^{(1)}(y_{\text{reh}})  + \beta_+ r^\ast H_{\nu_{II}}^{(2)}(y_{\text{reh}}) \right)\\
    &= - \frac{i \beta_-}{2} e^{\frac{i\pi}{4}(1+2 \nu_{II})}   \frac{\Gamma(3/2)}{\Gamma(3/2-1/p)} \left( \frac{y_{\text{reh}}}{2}\right)^{\nu_{II}-1/2}\\
    \gamma_-  &= \frac{H^2}{k^2}  \left( \frac{\eta_{\text{kin}}}{2\eta_{\text{reh}}}\right)^{1/p} 
\end{align}
Similarly,
\begin{equation}
    \gamma_+  = -\frac{H^2}{k^2}  \left( \frac{\eta_{\text{kin}}}{2\eta_{\text{reh}}}\right)^{1/p} 
\end{equation}



\bibliographystyle{JHEP}
\bibliography{ref.bib}

\end{document}